\documentclass[pdflatex,sn-nature]{sn-jnl}

\usepackage{graphicx}%
\usepackage{multirow}%
\usepackage{amsmath,amssymb,amsfonts}%
\usepackage{xcolor}%
\usepackage{textcomp}%
\usepackage{booktabs}%
\usepackage{rotating}%

\providecommand{\aap}{Astron. Astrophys.}

\providecommand{\apjl}{Astrophys. J. Lett.}
\providecommand{\pasp}{Publ. Astron. Soc. Pac.}
\providecommand{\mnras}{Mon. Not. R. Astron. Soc.}

\begin{document}

\title{Systematic non-compliance of satellite megaconstellations with IAU brightness limits}

\author[1,2]{\fnm{Jack} \sur{Rayworth}}\email{jjohn525@uwo.ca}

\author*[3,4]{\fnm{Denis} \sur{Vida}}\email{dvida@uwo.ca}

\author[1,3,4]{\fnm{Pauline} \sur{Barmby}}\email{pbarmby@uwo.ca}

\author[3,4]{\fnm{Peter G.} \sur{Brown}}\email{pbrown@uwo.ca}

\author[3,4]{\fnm{Sarah} \sur{Gallagher}}\email{sgalla4@uwo.ca}

\author[3]{\fnm{Michael J.} \sur{Mazur}}\email{mmazur5@uwo.ca}

\author[3,4]{\fnm{David L.} \sur{Clark}}\email{dclark56@uwo.ca}

\author[3]{\fnm{Maximilian} \sur{Vovk}}\email{mvovk@uwo.ca}

\author[3]{\fnm{Stanimir} \sur{Metchev}}\email{smetchev@uwo.ca}

\author[5]{\fnm{Lauchie} \sur{Scott}}\email{Lauchie.Scott@drdc-rddc.gc.ca}

\affil[1]{\orgname{Rotman Institute of Philosophy, The University of
Western Ontario}, \orgaddress{\city{London}, \state{Ontario},
\country{Canada}}}

\affil[2]{\orgname{Department of Philosophy, University of Oxford},
\orgaddress{\city{Oxford}, \country{United Kingdom}}}

\affil[3]{\orgname{Department of Physics and Astronomy, The University
of Western Ontario}, \orgaddress{\city{London}, \state{Ontario},
\country{Canada}}}

\affil[4]{\orgname{Institute for Earth and Space Exploration, The
University of Western Ontario}, \orgaddress{\city{London},
\state{Ontario}, \country{Canada}}}

\affil[5]{\orgname{Defence R\&D Canada Ottawa},
\orgaddress{\city{Ottawa}, \state{Ontario}, \country{Canada}}}

\abstract{Satellite megaconstellations in low Earth orbit interfere with
ground-based optical astronomy, but the scale of that interference has
been hard to measure. Here we report an uncued
photometric survey from a mid-northern latitude site in late 2024 which
recorded some 25,000 tracks of 5,118 individual satellites to a limiting
visual magnitude of $+9.0$. The observed population is systematically
brighter than the altitude-adjusted limit of order seventh magnitude
recommended by the International Astronomical Union Centre for the Protection of the Dark and Quiet Sky from Satellite Constellation Interference (IAU CPS). Starlink satellites
exceeded that limit on roughly a quarter of all geometrically possible
passes, rocket bodies on about a fifth, and the emerging
Guowang/Hulianwang constellation on more than half. OneWeb was the most compliant major operator (exceeding the IAU CPS limit on only
about one in twenty possible passes), which shows
that effective brightness mitigation is achievable with current
technology. Satellite visibility is concentrated into the two hours after sunset
and before sunrise and into limited regions of the sky. From the
measured plane-of-sky satellite flux we predict that of order one in
three 30~s twilight exposures at the Vera C. Rubin Observatory will
carry a satellite trail.}

\maketitle

Humans have been launching satellites into space for nearly 70 years,
but the number in orbit has risen sharply over the last few years. That
rise is driven almost entirely by constellations such as Starlink,
OneWeb, BlueBird, Qianfan and Guowang \citep{JSR_2025}. These
constellations carry substantial benefits (communications to remote
areas and remote monitoring of the Earth among them) alongside
substantial harms, which include interference with astronomical
observations both through the direct appearance of trails in images
\citep{hainaut_impact_2020} and through an increase in the diffuse sky
background \citep{barentine_noise_2023}, as well as the production of
orbiting debris \citep{boley_satellite_2021}.

Within a year of the
first Starlink launches in 2019 \citep{hainaut_impact_2020,
mcdowell_low_2020, williams_eso_2021} the {\sc satcon 1} conference had
produced a report on the impact of satellite brightness together with a set of recommendations
for mitigation \citep{walker_satcon1_2020}. Recommendation 5 specified
that the brightness of a satellite in reflected sunlight should meet an
altitude-adjusted limit,

\begin{equation}
V > \max\left[\,7.0,\ 7.0 + 2.5\log_{10}(A/550\ \mathrm{km})\,\right],
\label{eq:iau}
\end{equation}

\noindent where $V$ is the apparent magnitude in the Johnson--Cousins
$V$ band and $A$ is the orbital altitude: that is, magnitude 7.0 at or
below 550~km and the altitude-scaled value above it, whichever is the
fainter. The altitude dependence
matters because, although satellites in higher orbits are fainter, their
slower angular velocities raise the surface brightness of the trail left
in a sidereally tracked image. Note that the limit does not continue to
brighten below 550~km, since there the constraint is set by detector
saturation rather than by trail surface brightness. This recommendation was affirmed by the International
Astronomical Union's Centre for the Protection of the Dark and Quiet Sky
from Satellite Constellation Interference (IAU CPS)
\citep{iau_cps_2024}, with further explanation given by
\citep{boley_iau_2025}. Satellite constellations also affect radio
astronomy \citep{grigg_growing_2025}, which we do not consider here.

Since 2020 a range of modelling studies have predicted satellite
brightness and sky distribution as a function of location, season and
solar time \citep{lawler_visibility_2022, hall_semi-empirical_2022,
bassa_analytical_2022}. Others have gone on to quantify the consequences
for spectroscopy \citep{hainaut_spect_2024}, for wide-field surveys such
as the Vera C. Rubin Observatory's Legacy Survey of Space and Time
\citep{kandula_simulated_2025, hasan_mitigating_2022, hu_avoidance_2022}
and even for space-based telescopes (about 3\% of Hubble Space Telescope
images taken before 2021 already show satellite trails)
\citep{kruk_hubble_2023, borlaff2025satellite}. All of them depend on
empirical satellite photometry for calibration, and such photometry
remains scarce. Note that the properties which make satellites harmful
to astronomy (rapid angular motion, brightness, and a sky position which
changes as orbits are adjusted) are the same properties which make them
challenges to observe. Campaigns to date have therefore
either tracked or ``snapshotted'' modest numbers of pre-selected targets
\citep{krantz_steward_2023, zhi_multicolor_2024, caddy_daytime_2025}.
The alternative has been visual estimates together with limited open source instrumental observations such as the
Mini-MegaTORTORA database \citep{karpov_mmt_2016,
mallama_satellite_2025}. That database is extensive, but it does not
contain a complete satellite listing. We discuss this earlier work in Supplementary
Note 1.

Here we fill the gap between well-documented but targeted studies and
less well-documented survey data, using the Luciole sensor system
\citep{vida2024project}. Luciole draws on hardware and software heritage
developed for meteor studies (the Global Meteor Network) and later
adapted to satellite tracking.
Rather than pointing at known satellite positions, it casts an optical
``net'' over most of the sky and records at 25 samples per second. The result is an uncued, photometrically
calibrated census of the low Earth orbit (LEO) population as it appears
from the ground, from which we derive a compliance assessment for
individual constellation brightness together with an instrument-independent
measure of how often satellites will contaminate a given astronomical
observation.

\section{Results}

\subsection{Brightness of the observed satellite population}

During a 36-night campaign from 7 October to 11 November 2024 at
Elginfield Observatory, Ontario ($43.19^{\circ}$N, $81.32^{\circ}$W),
six narrow-field Luciole cameras recorded, calibrated and analysed
24,812 satellite tracks from 5,118 unique satellites in 230 clear hours
drawn from 407 hours of possible observing time (Methods; Extended Data
Figs.~\ref{edfig:system} and \ref{edfig:fov}). The result is a month-long snapshot
of the LEO environment visible from a mid-northern latitude site as of
late 2024.

\begin{figure}[p]
\centering
\includegraphics[width=\linewidth]{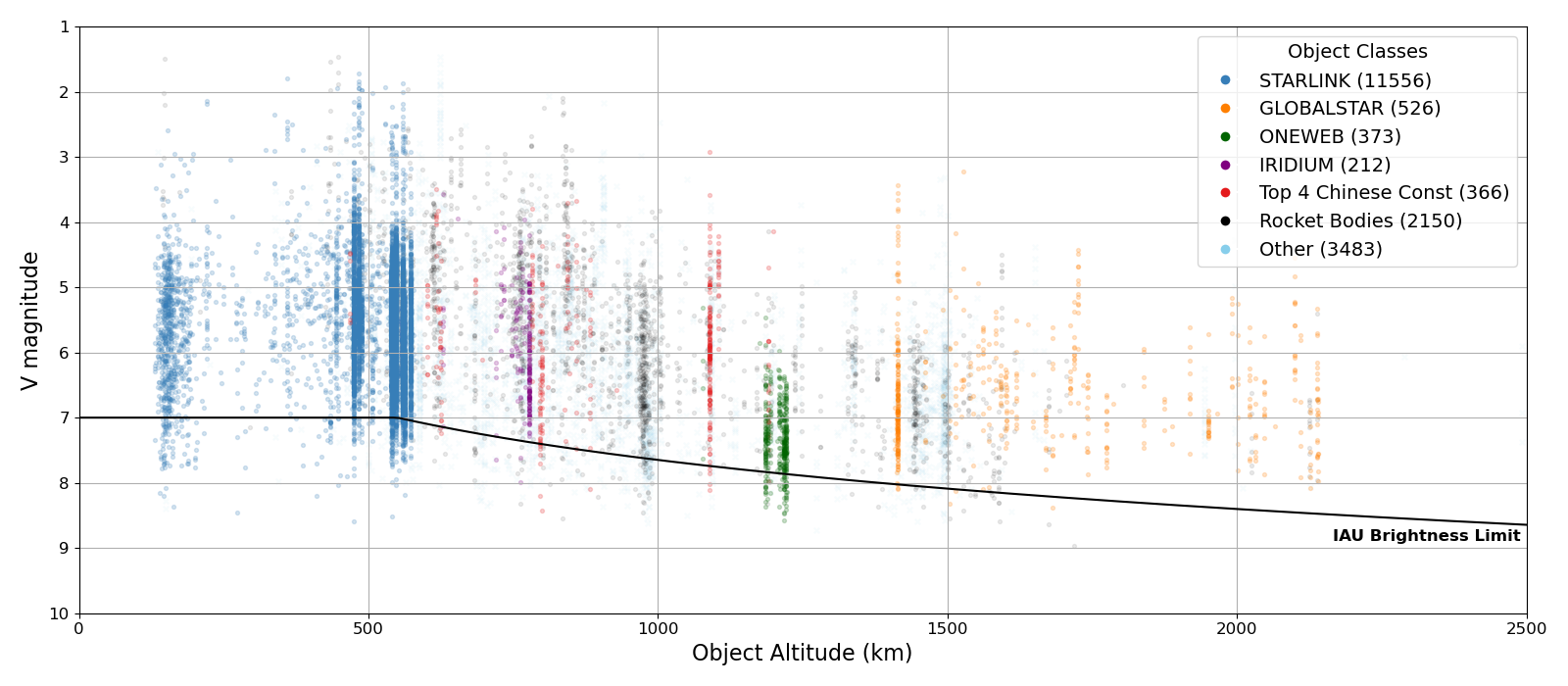}
\caption{\textbf{Brightness of the observed satellite population as a
function of orbital altitude.} Each point is one tracklet, plotted at the
median visual magnitude of its track points and the orbital altitude of
the observed object. Six satellite constellations and object classes are
colourised; all remaining objects are shown in light blue. The black line is
the altitude-adjusted IAU CPS brightness limit of
equation~(\ref{eq:iau}). Objects above the line in this orientation
(brighter magnitudes) exceed the recommended limit. The concentration of
non-compliant objects between 500 and 600~km corresponds to the dense
Starlink shell.}
\label{fig:mag_scatter}
\end{figure}

\begin{figure}[p]
\centering
\includegraphics[width=\linewidth]{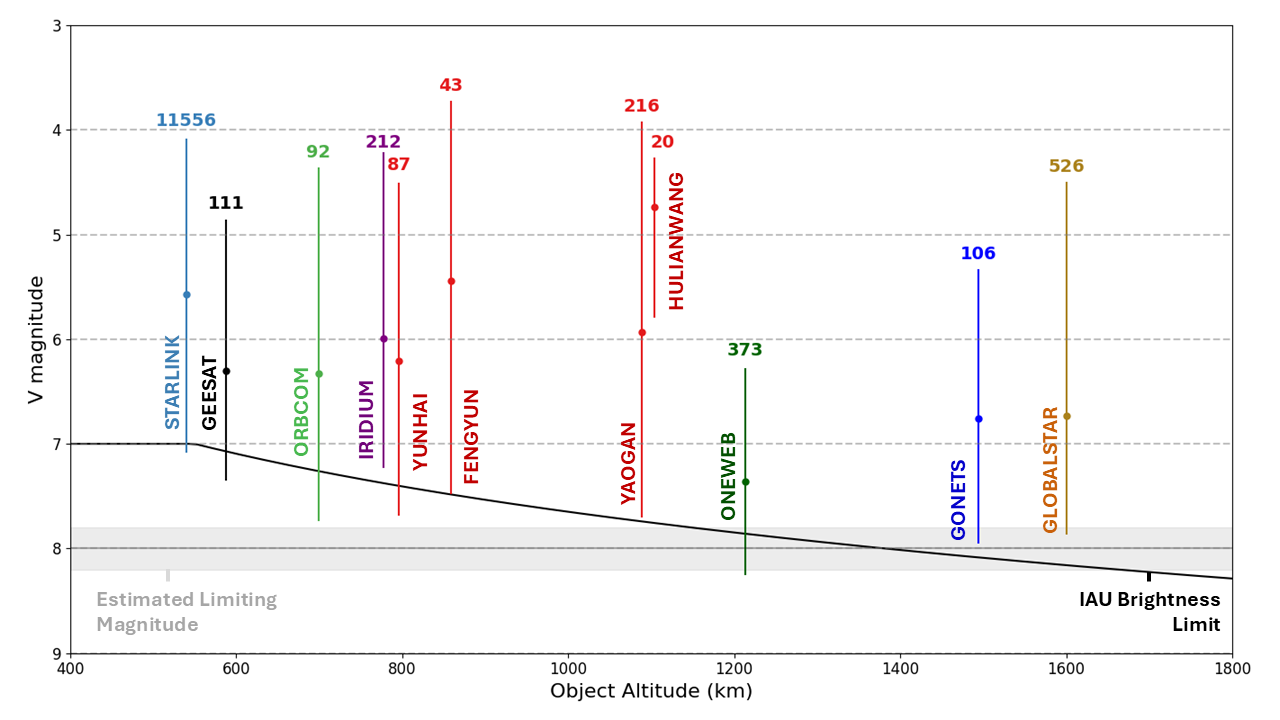}
\caption{\textbf{Observed brightness of the most frequently detected
constellations.} Dots give the median visual magnitude of each
constellation as recorded in our observations, positioned at the median
altitude of the observed objects of that constellation; bars span the
2.5\%--97.5\% percentile interval (the central 95\%) of the visual magnitude. The number of
observed tracks in each constellation is shown above each bar. The black
line is the IAU CPS brightness limit. OneWeb is substantially fainter
than the other major constellations, and this cannot be attributed to
its higher orbit alone, since Gonets and Globalstar lie at comparable or
greater altitudes and appear substantially brighter.}
\label{fig:errbars}
\end{figure}

Figure~\ref{fig:mag_scatter} shows the measured apparent brightness of
every track as a function of orbital altitude. A large fraction of
the observed satellites lie brighter than the limit recommended by the IAU
CPS (equation~\ref{eq:iau}).
This is most pronounced in the dense orbital shell between 500 and
600~km, which is predominantly occupied by Starlink and which, through
sheer numbers, dominates our dataset (the same concentration is shown as
a density map in Extended Data Fig.~\ref{edfig:density}).
Figure~\ref{fig:errbars} compares the median and the 2.5\%--97.5\%
percentile interval of the brightness for the eight most frequently
observed constellations together with rocket bodies and other objects. Extended Data
Table~\ref{edtab:constellations} gives summary statistics for the 25 groups with
the most observed passes brighter than the limit, together with Fengyun and
Guowang/Hulianwang.

\begin{figure}[p]
\centering
\includegraphics[width=\linewidth]{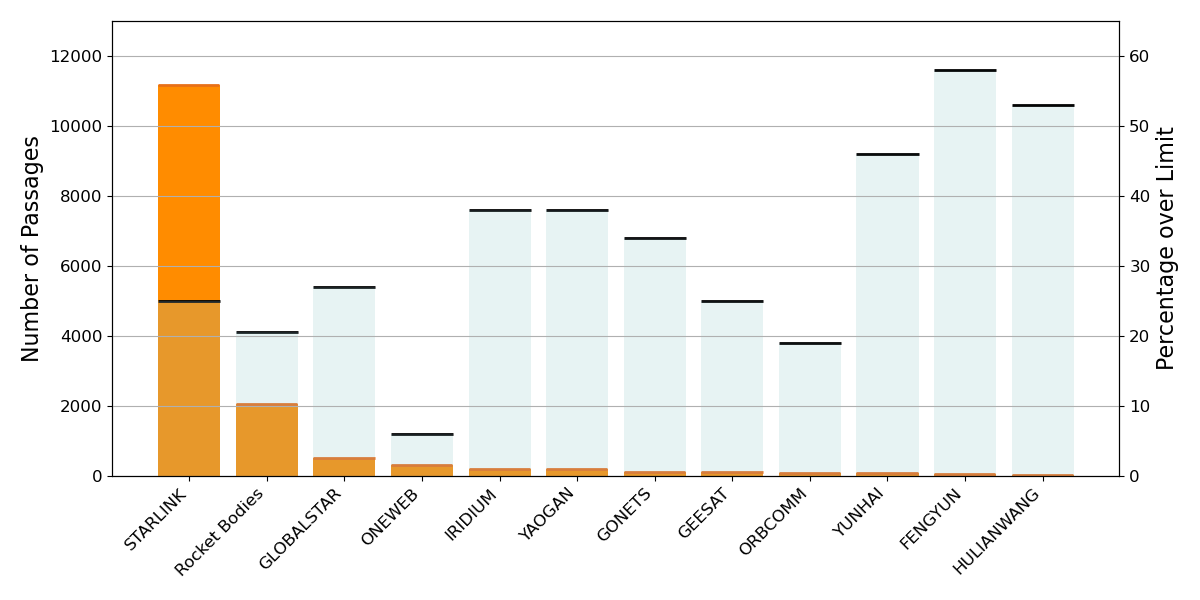}
\caption{\textbf{Non-compliance with the IAU CPS brightness limit by
object class.} Orange bars (left axis) give the absolute number of
observed tracks brighter than the altitude-adjusted IAU limit,
categorised by satellite constellation or, for debris, by launch-vehicle
origin. Light blue bars (right axis) give the same quantity as a
percentage of all theoretically possible observations of that class,
that is, instances in which an illuminated object from the class entered
the field of view of an operating camera. All categories shown had at least 18 observed passages. The percentage is a rough indicator of
the intrinsic brightness of the class and is independent of constellation
size, so it would not be expected to change were a constellation scaled
up in number; it therefore provides a way to predict the total impact of
a planned expansion.}
\label{fig:compliance}
\end{figure}

Starlink produced the highest absolute number of tracks exceeding the threshold, making up roughly 2/3 of all observed passes brighter than the limit
(Fig.~\ref{fig:compliance}). However, the absolute count is dominated by
constellation size. A more physically meaningful comparison is found by expressing
non-compliant passes as a percentage of the total number of theoretically possible
observations. We define these as those instances where public two-line elements
predict an illuminated object entering the field of view of an operating
camera (after accounting for cloud, camera downtime and loss through
atmospheric attenuation). 

Using this measure Starlink exceeded the limit on
$\sim25\%$ of possible passes and rocket bodies on $\sim21\%$, while the
Guowang/Hulianwang and Fengyun groups exceeded the IAU CPS limit on more than 50\%. The Guowang/Hulianwang figure is the
more concerning of these, given plans to expand that constellation to 13,000
satellites.\footnote{SpaceNews,
\url{https://spacenews.com/china-is-developing-plans-for-a-13000-satellite-communications-megaconstellation/},
accessed 30 September 2025.} OneWeb stands alone among the major
constellations with a minimal brightness exceedance of only $\sim6\%$ (median $V = 7.3$, with a 2.5\%--97.5\%
percentile interval of 6.3--8.3). Note that this percentage is independent of
constellation size (it would not be expected to change were a
constellation scaled up), so it also provides a way to predict the
impact of a planned expansion.

\subsection{Observing geometry and the twilight window}

\begin{figure}[p]
\centering
\includegraphics[width=\linewidth,trim={0cm 0 0cm 0},clip]{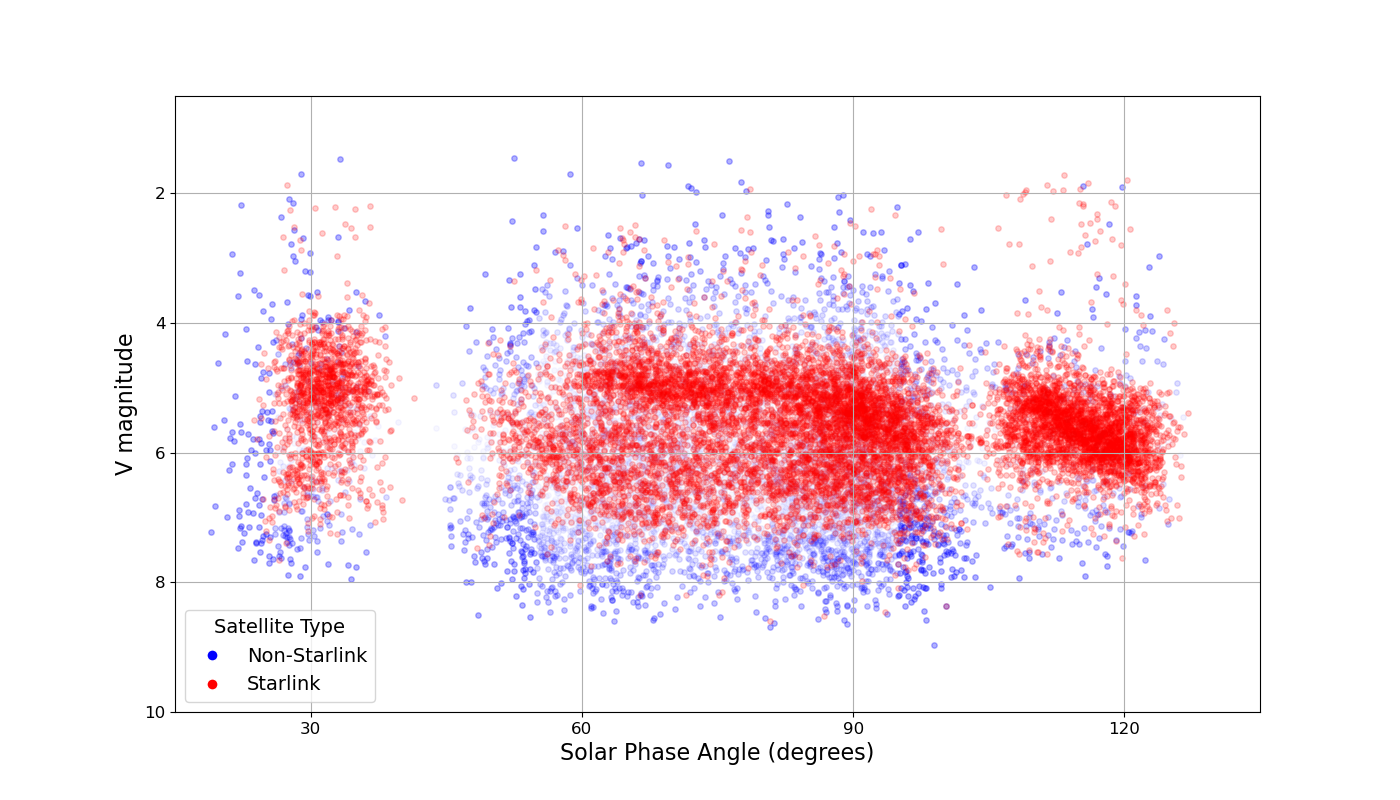}
\caption{\textbf{Solar phase angle against visual magnitude for observed
tracks.} Phase angle is evaluated at the midpoint of each track (maximum
track length $\sim$20$^{\circ}$) and visual magnitude is the median value
exhibited during the track. Starlink and non-Starlink objects are shown
separately; to make the Starlink structures visible, non-Starlink track
points are rendered more transparent around them. The three groupings
correspond to solar array backscatter ($\sim$30$^{\circ}$), diffuse
scattering from the satellite chassis ($\sim$60$^{\circ}$--100$^{\circ}$)
and specular reflection from flat surfaces ($\sim$115$^{\circ}$). The
gaps between the groupings are an artefact of the fixed narrow-field
camera pointing and are absent from the all-sky wide-field data.}
\label{fig:phase}
\end{figure}

The apparent magnitude of a satellite depends strongly on the solar
phase angle, that is on the angle between the observer, the satellite
and the Sun. Figure~\ref{fig:phase} shows this relationship for all
observed tracks. The distribution is not uniform but breaks into three
distinct clusters. The gaps between them are an observational artefact:
satellite observations from Luciole's less sensitive but all-sky
wide-field system ($+8$ mag) show no such clustering, a direct indication
that the gaps arise from the limited pointing geometry of the
narrow-field cameras. The three groupings themselves nonetheless align
with the three known scattering regimes: a population at low phase
angles ($\sim30^{\circ}$) likely corresponds to solar array backscatter,
a broad central cluster ($\sim60^{\circ}$--$100^{\circ}$) to diffuse
scattering from the satellite chassis, and a more concentrated cluster
at high phase angles ($\sim115^{\circ}$) to specular reflection from
flat surfaces \citep{caddy2024surprising}. The pattern is present in
both the Starlink and the non-Starlink populations. Starlink satellites
are far more strongly concentrated at the specular phase angles
and form a brighter upper envelope in the diffuse regime.

\begin{figure}[p]
\centering
\includegraphics[width=\linewidth,trim={2cm 0 2.5cm 0cm},clip]{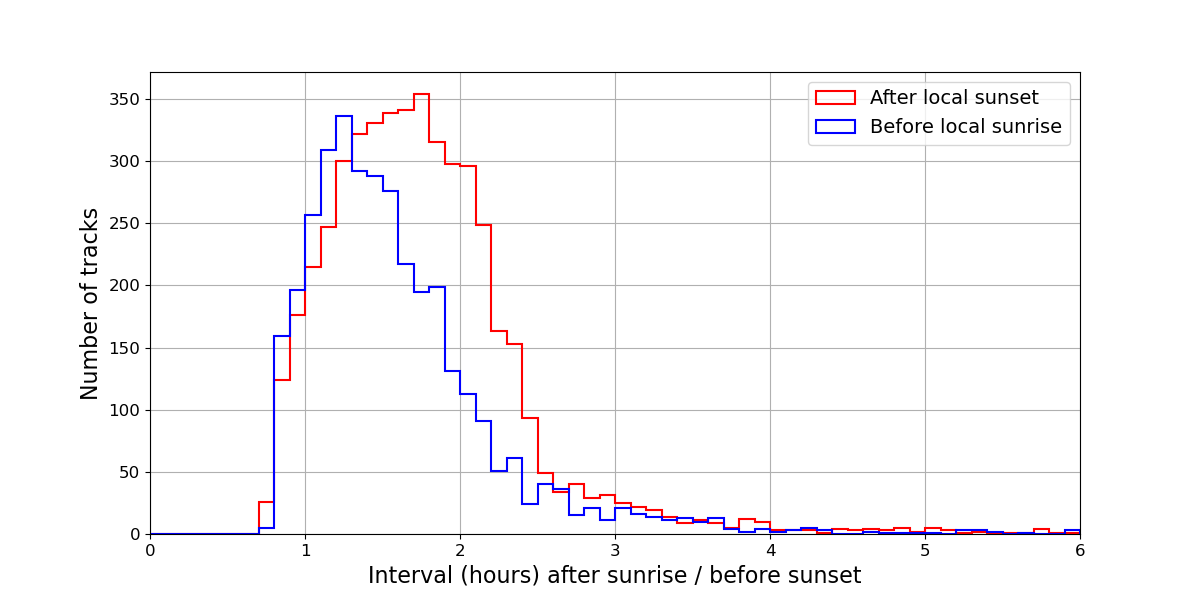}\\[2ex]
\includegraphics[width=\linewidth]{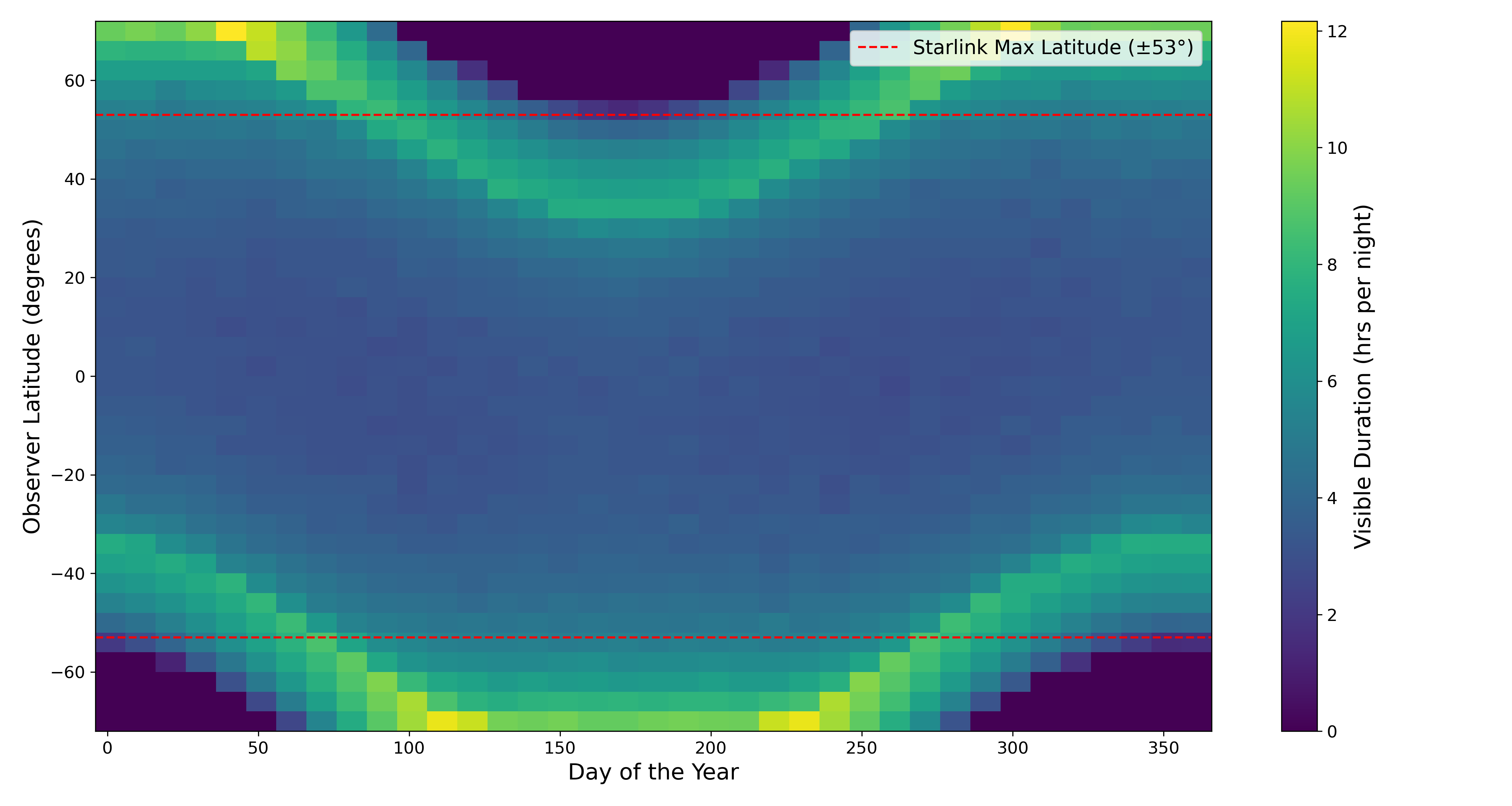}
\caption{\textbf{Concentration of satellite visibility into the twilight
window.} \textbf{a}, Histogram of track detection times measured relative
to local sunset (red) and sunrise (blue) from Elginfield Observatory in
October--November 2024. The asymmetry arises from asymmetric camera
coverage: no camera in the narrow-field survey pointed east.
\textbf{b}, Modelled duration of satellite visibility through the year as
a function of observer latitude, for a 550~km shell. Horizontal red lines
mark the latitudinal extent of the Starlink megaconstellation, which
dominates the absolute numbers of visible satellites. Visibility duration
is nearly invariant through the year between $-35^{\circ}$ and
$+35^{\circ}$ latitude at about 3.33~hours per night.}
\label{fig:timing}
\end{figure}

Satellite visibility is constrained almost entirely to twilight
(Fig.~\ref{fig:timing}a). The vast majority of tracks were recorded
within the first two hours after sunset and the last two hours before
sunrise. Through the middle of the night most LEO satellites are in the
Earth's shadow and are therefore unilluminated, in accord with existing
model predictions \citep[e.g.][]{muirhead2025modeling}. The slight
asymmetry in our detections (more after sunset than before sunrise) is
an artefact of a camera configuration which had more coverage of the
western sky.

To generalise this result beyond our own site we developed a geometric
visibility model. The model treats a constellation as a spherical shell
at a fixed altitude of 550~km and asks, at each time step, whether any
part of the shell is simultaneously above the observer's horizon,
sunlit, within the observed phase angle range of
$25.9^{\circ}$--$123.2^{\circ}$ and seen against a dark sky (Methods;
Extended Data Fig.~\ref{edfig:funnel}). Applied globally
(Fig.~\ref{fig:timing}b), it gives the number of hours where satellites are observable. This is
found to be essentially constant through the year for latitudes between
$-35^{\circ}$ and $+35^{\circ}$, at 3.33 hours (2.5\%--97.5\% percentile
interval 2.92--5.70 hours). In effect there are of order 1.5--2 hours of
intense megaconstellation visibility after sunset and as long again before
sunrise, every day of the year, at the latitudes where most of the
world's population lives. Note that this applies to shells near 550~km. The
same model gives 2--3.6 hours for shells at 350--480~km but 7--10 hours, most of
the night, at 1,200~km (Methods), so constellations in higher orbits such as
OneWeb and Guowang are not confined to twilight. Expressed instead as a fraction of the night
(Extended Data Fig.~\ref{edfig:vispct}), the largest impact falls in
summer at high latitudes, where satellites remain illuminated for the
whole night, though the absolute time lost is then small because the
nights themselves are short. Even so, satellites occupy only part of the sky:
for the 550~km shell viewed from our Elginfield survey station they are visible over about
60\% of the sky just after sunset, dropping to 35\% halfway through the
visibility window. 

\subsection{Plane-of-sky satellite flux and predicted contamination}

To express the passage rate in a form independent of our instrumentation
we define the plane-of-sky satellite flux $F_{sat}$ as the number of
satellite passages per square degree per unit time (Methods). Averaged
over the six cameras and over the twilight visibility periods, the flux
for all objects brighter than our effective completeness limit of
$+6.15$ mag is $(3.51 \pm 1.5) \times 10^{-3}$ satellites deg$^{-2}$
min$^{-1}$. Correcting this measurement to the IAU CPS limit with the
piecewise power law of equation~(\ref{eq:power-law}) gives $(5.56 \pm 2.4) \times 10^{-3}$ satellites deg$^{-2}$ min$^{-1}$ brighter than $+7.0$ mag. This exceeds the flux quoted above because it is a completeness-corrected population estimate and not a raw count (14,997 of the 15,591 recorded passages were brighter than the altitude-adjusted IAU CPS limit, and the survey is complete only to $+6.15$). The
Elginfield survey per-camera values span a factor of $\sim$3 (Extended Data
Table~\ref{edtab:flux}), a direct consequence of their different
pointing directions. They show how anisotropic the satellite
distribution on the sky actually is.

The flux converts into the number of satellites expected in an arbitrary
observation once it is combined with the field of view and an effective
collection time $T_{eff} = t_{exp} + t_{pass}$, where $t_{pass}$ is the
mean time a satellite takes to cross the field (Methods). A single
expression then spans both regimes: the short-exposure ``snapshot''
regime, where the count is set by the instantaneous satellite density in
the field, and the long-exposure ``flux'' regime, where it is dominated
by new satellites entering during an on-sky exposure.

Scaling to an arbitrary
limiting magnitude requires the satellite magnitude distribution, which we
measure directly to our effective completeness limit of $+6.15$ mag and extend to
fainter magnitudes using catalogued radar cross-sections (Methods;
Extended Data Fig.~\ref{edfig:distributions}), giving population indices (ratio of the number of satellites of brightness $M+1$ to those of brightness $M$)
of $r = 4.62$ measured between $+2.5$ and $+5.2$, the fit from which the completeness limit above is derived. At fainter magnitudes we extrapolate the satellite population using published Radar Cross Section values and find the population index falls to $r = 1.72$ (Methods).
The latter value should be considered the more uncertain of the two
as it assumes an RCS-to-magnitude conversion rather than direct
photometry.

Table~\ref{tab:predictions} applies this framework to four representative
observing scenarios. A human observer under a dark sky ($+6.5$ limiting
magnitude) would see of order 44 bright satellites at any instant within
their field of view during the twilight window. Across the whole sky the
figure rises to of order 140. For the Vera C. Rubin Observatory,
trailing loss for an
object moving at the median observed angular speed of
$0.56^{\circ}$~s$^{-1}$ reduces the effective limiting magnitude for
satellite detection from $+24$ to $+11.49$ (Methods), and scaling the
measured flux to that limit gives an enhancement factor of 11.4 relative
to the IAU limit. Rubin should therefore find $\sim0.34$ satellite trails
in every 30~s exposure taken during the intensive visibility period,
which comprises $3.58^{+2.10}_{-0.26}$ hours of the night through the
year at the latitude of Rubin (median and 2.5\%--97.5\% range) (consistent with the independent estimate of
\citep{hu_avoidance_2022}). A consumer APS-C camera with a 20~mm lens sits
at the opposite extreme. A trailing loss of 5.3 mag leaves it an
effective satellite limiting magnitude of $+2.71$ and a predicted rate
of only $\sim0.054$ trails per image (so that only exceptionally bright
passes are recorded).

\section{Discussion}

The principal finding of this survey is that a large fraction of the
present LEO satellite population is systematically brighter than the IAU
CPS limit. This is not a problem confined to occasional anomalously
bright glints (which we do not model), but a baseline characteristic of the current orbital
environment. The median observed magnitude for all of the major
constellations (Starlink, Yaogan, Fengyun and Globalstar among them) is
substantially brighter than the corresponding IAU limit at their
operational altitude indicating that a large proportion of passages from these constellations
may contaminate optical measurements. That shifts the focus of
mitigation away from preventing rare bright glints and towards the
baseline reflectivity of the bus, the antennas and the solar arrays
(characteristics fixed at the design and manufacturing stage rather than
in operations). 

From our survey we find that OneWeb is substantially dimmer than the other
main constellations.  Fig.~\ref{fig:errbars} makes clear that this
is not attributed to its higher (1,200~km) orbit alone, since Gonets
and Globalstar lie at comparable or greater altitudes and appear
substantially brighter. OneWeb also has the largest fraction of
observed passes fainter than the IAU limit of any major constellation (15\%, against
1--11\% for the others; Extended Data Table~\ref{edtab:constellations}). That
is in general accord with other recent comprehensive studies
\citep{mallama_satellite_2025, krantz_steward_2023}. Our median
magnitudes agree with those of \citep{mallama_satellite_2025} to within
$0.05$~mag for the classes our survey samples completely, and are
brighter by $0.55$--$0.93$~mag for those lying fainter than our
completeness limit of $+6.15$, as expected from magnitude-dependent
detection efficiency. Our compliance percentages for such classes
(OneWeb among them) should therefore be read as lower limits. The full
comparison is given in Supplementary Note 3.

Starlink dominates in the absolute number of satellites brighter than the IAU limit. 
This is true even though only about one in four
possible Starlink passages exceeds the limit; the sheer size of the
constellation causes it to dominate. 

Most concerning are our findings for the Guowang/Hulianwang satellites, which were brighter than the limit in 100\% of observed passes (and in more than
half of geometrically possible ones). If the plan to launch 13,000 of
them is realised, they will become the dominant source of optical sky
pollution unless mitigation is implemented. 

Growth in large debris is a side effect of the increased satellite population. BREEZE and Long March rocket bodies were non-compliant in 99\% and 97\% of
observed passages respectively (Extended Data
Table~\ref{edtab:constellations}). Any comprehensive solution must
therefore address the growing population of spent upper stages and larger mission-related debris as well
as the well-known megaconstellations.

The phase angle behaviour found in our survey points to a physical cause of these results. The strong
Starlink concentration in the specular scattering peak near $115^{\circ}$, together
with its brighter upper envelope in the diffuse regime, implies that the
Starlink design is inherently more prone to bright specular glints than
the average LEO object. The large flat phased-array antennas are the
plausible cause (though we cannot separate that from attitude effects
observationally), and if so then mitigation must go beyond
darkening surfaces (which acts mainly on the diffuse component) and must
also manage large flat surfaces so as to suppress the specular one. We
note that brightness has not in fact improved monotonically across
generations. The Version 2.0 Mini Direct-to-Cell satellites reverse the
dimming trend established between 2020 and early 2024, returning close
to the brightness of the original Version 1.0 spacecraft (Extended Data
Fig.~\ref{edfig:starlinkgens}; Supplementary Note 5).

Our survey and associated satellite brightness model show that the operational cost to astronomy is set by timing as much as by
brightness. Concentration of visibility into the twilight hours creates
a ``twilight bottleneck''. This is because the period of greatest satellite
visibility coincides with the period of great value to ground-based
astronomy as unique data can only be obtained in this interval. For example, observers use these hours to obtain sky flats and other
required calibration data \citep{tyson1993exposure} (which satellite
trails can render useless, degrading all the science data which depend
on them). Several science programmes are themselves confined to
twilight, most notably the discovery and characterisation of near-Earth
objects and other inner Solar System bodies (best observed at their
largest solar elongation) \citep{ye2020twilight, sheppard2022deep}. Our
visibility model shows that this is neither localised nor seasonal:
mid-latitude sites, where most of the world's premier research
facilities are located, face a persistent year-round window of intense
satellite visibility of $\sim$3 hours each night from the population below
$\sim$600~km. Satellites may be visible
for only a fraction of a long winter night, but they contaminate nearly
100\% of the twilight window itself. The impact is thus not a
statistical loss of observing time so much as a targeted threat to
particular scientific goals (planetary defence among them).

Expressing our results as a standardised flux makes them quantitatively easy to apply to other sites and instruments.
The $F_{sat}$ metric is a fundamental measure of on-sky satellite traffic density which
any observatory can apply (in units of satellites deg$^{-2}$
min$^{-1}$). It allows direct comparison between surveys
and provides a simple framework for assessing proposed constellations.
Because it is based on empirical measurement it can be updated as
traffic conditions evolve (which we expect will be measured on an ongoing basis with Project
Luciole). As applied to the Rubin Observatory, we
predict a contaminating trail in of order one out of every three
exposures taken during the twilight period, a substantial rate of data
loss which quantifies concerns the community has raised since the first
megaconstellation deployments \citep{hainaut_impact_2020,
walker_satcon1_2020, mcdowell_low_2020, tysonMitigationLEOSatellite2020}.

Our survey and model has several limitations. The data come from a single mid-northern latitude site over
a finite 36-night period, so that while the visibility model generalises
the timing globally, the measured flux is specific to our location and
epoch (and the LEO population is in any case highly dynamic). The fixed camera
pointing introduced selection effects (evident in the gaps in phase
angle coverage of Fig.~\ref{fig:phase}) which an all-sky monitoring
system would remove. Our absolute detection efficiency falls fainter than the effective
completeness limit of $+6.15$, so that the median magnitudes of classes lying
mostly fainter than that limit (OneWeb among them) are biased bright and their
compliance percentages are lower limits (Supplementary Note 3). The altitudes
used to evaluate the IAU CPS limit are catalogue values from June 2025 rather
than at the epoch of observation, which matters only for objects that changed
orbit between the two. Conversion from instrumental to $V$-band magnitude
rests on a synthetic transformation built from an assumed average
satellite spectrum (which yields a colour index of $-0.02$ mag;
Methods), and multi-band photometry will be needed to
constrain surface albedo and composition to build realistic
bidirectional reflectance distribution functions
\citep{fankhauser2023satellite, littoriano2024modelling}. 

Finally, the
survey does not generally capture satellites undergoing post-launch
orbit raising (public two-line elements are usually unavailable
immediately post-launch and through the first significant orbit-raising
interval \citep{Kuciapinski2023}). This is significant as orbit raising is a major source of high-luminosity events, as
satellites often adopt non-nominal attitudes and are at lower altitudes hence substantially
brighter than when on station \citep{mallama2024brightness,
Mallama2024DTC}. Our flux and magnitude distributions therefore sample
only the stable catalogued population, and should be considered a lower
limit on the true rate of the most disruptive events.

Overall, non-compliance with astronomical brightness standards is the
norm rather than the exception for the current LEO population, it is
concentrated in precisely the hours which matter most to ground-based
astronomy, and its observational cost can now be predicted
quantitatively for any instrument. The success of some operators in
producing fainter satellites provides a clear and actionable benchmark
for the rest of the industry. Future monitoring of this environmental issue requires continuous
monitoring from a geographically distributed sensor network, long-term photometric tracking of constellations from launch onwards (the only way
to assess how well mitigation measures actually work, and at the 25~Hz cadence
of Luciole able to resolve the rotation of tumbling objects; Extended Data
Fig.~\ref{edfig:cosmos}), and multi-colour photometry to build a library of
satellite reflectance functions.

\section{Methods}

\subsection{The Luciole sensor system}

Our data are produced by the Luciole sensor system, an array of
sky-facing cameras designed for satellite tracking and deployed across
four sites in Canada (Ontario, Saskatchewan, British Columbia and
Nunavut; Extended Data Fig.~\ref{edfig:system}). The system is a direct
adaptation of hardware and software originally developed for meteor
observations and now optimised for space situational awareness (SSA).
The technology derives from the Global Meteor Network (GMN), a worldwide
project that uses low-cost, high-cadence video cameras for continuous,
wide-field monitoring of the night sky \citep{vida2021global}. GMN
employs commercial-off-the-shelf (COTS) video cameras at a frame rate of
25~Hz, with software optimised for detecting transient linear phenomena.
Satellites may be viewed as very low velocity meteors of long duration,
so adapting the system was straightforward, and a decade of operational
GMN experience provides a well-tested and validated method for the new
application.

The Luciole hardware consists of a ``fly's eye'' array of 14 COTS
low-light video cameras ($1920 \times 1080$ pixels) equipped with Sony
STARVIS2 sensors, covering the entire sky above $30^{\circ}$ elevation
from a single site. The array comprises two sub-systems: eight
wide-field cameras with 8~mm f/1.0 lenses giving a field of view (FOV)
of $57^{\circ} \times 30^{\circ}$ and a plate scale of $\sim$1.7~arcmin per pixel, and six more sensitive narrow-field cameras with 25~mm f/1.0
lenses, a FOV of $17^{\circ} \times 10^{\circ}$ and a plate scale of
$\sim$0.5~arcmin per pixel \citep{vida2024project}. The total FOV of the
entire system is 14,700~deg$^{2}$, mainly provided by the wide-field
cameras. The narrow-field cameras have a limiting $V$-band magnitude of
around $+9.0$ for satellites, corresponding to $\sim$1~m objects for an
assumed Lambertian sphere scattering at a $90^{\circ}$ solar phase angle
with an albedo of 10\% at an altitude of 1,000~km \citep{vida2024project}.

The detection algorithm, originally designed to identify the linear
streaks of meteors, was modified to track the near-linear motion of
satellites across consecutive video frames. A key advantage of the 25~Hz
cadence is that most LEO satellites are captured as point sources rather
than streaks, which simplifies centroiding: a typical Starlink satellite
moving at $\sim$0.5~deg~s$^{-1}$ traverses just over 1~arcmin during a
single frame exposure, less than one pixel for the 8~mm systems and
approaching two pixels for the 25~mm systems, whereas stars move at most
0.01~arcmin per frame. The software operates in a continuous, uncued
staring mode, detecting all moving objects that cross the field.
Detections are correlated against the public satellite catalogue from
Space-Track.org\footnote{\url{https://www.space-track.org/}, accessed 6
October 2025.} using a novel algorithm that achieves correlation at the
80--90\% level on clear, moonless nights. Throughout 2024 to mid-2025,
Luciole autonomously tracked around 20,000 unique catalogued objects
from the four Canadian sites. Passage simulations indicate that
approximately 30\% of satellite transits over a given site intersect the
field of view of at least one narrow-field camera.

\subsection{Astrometry, photometry and timing}

\textit{Astrometry.} The system provides continuous angular position
measurements throughout a satellite's passage. The narrow-field cameras
achieve a single-frame astrometric accuracy of approximately
5~arcseconds and the wide-field cameras $\sim$13~arcseconds; these
figures represent the astrometric plate residuals across the field
following the procedure of \citep{vida2021global}. This precision is
made possible by an astrometric calibration method that accurately
models the significant optical distortions present in wide-field lenses
\citep{vida2021global}. As demonstrated by \citep{vida2024project}, the
high data rate allows orbit determinations that match or exceed the
accuracy of traditional, higher-precision but lower-cadence sensors. A
smoothing algorithm based on the physical motion of satellites reduces
measurement uncertainties to 1--2~arcseconds with full accounting of the
associated uncertainties.

\textit{Photometry.} The system measures apparent brightness in the
wide-band (unfiltered) instrumental bandpass with a precision of
$\pm0.15$~mag, representing the formal confidence interval of the
photometric zero point of the stellar fit as described in
\citep{vida2021global}, found using a 256-frame average stack.
Photometric calibration is performed for each track using background
stars and a synthetic band constructed to match the spectral sensitivity
of the cameras. A custom star catalogue combines Gaia DR3
\citep{vallenari2023gaia} positions and magnitudes with the Sky2000
catalogue for $U$, $B$, $V$, $R$, $I$ bands \citep{myers2001vizier}. The
method includes corrections for atmospheric extinction and lens
vignetting \citep{vida2021global}. The high-cadence photometry is
particularly valuable for lightcurves, which can be used to monitor
satellite health and stability by determining rotational period and
identifying uncontrolled tumbling; Extended Data
Fig.~\ref{edfig:cosmos} shows the 25~Hz lightcurve of the tumbling
object Cosmos 2553 as a demonstration.

\textit{Timing.} All measurements are timed with millisecond-level
accuracy. The system clock is conditioned using a GPS device and the
\texttt{chrony} network time protocol service.\footnote{\url{https://chrony-project.org/},
accessed 1 October 2025.} A fixed instrumental delay between frame
capture and data logging is calibrated and removed from the final data
products \citep{vida2024project} using a selection of satellites with
available high-accuracy ephemerides.

\subsection{Conversion of instrumental magnitude to the V band}

To allow comparison with other SSA datasets, a synthetic colour
transformation was derived to convert the instrumental magnitude of the
Luciole system to the standard Johnson--Cousins $V$ band. The spectral
sensitivity of the complete optical system was measured in the
laboratory using a calibrated monochromatic light source, giving the
instrumental bandpass $T_{inst}(\lambda)$. For application to satellites, we assume the illuminating source was
the standard AM1.5G solar spectrum $S(\lambda)$
\citep{gueymard2002proposed}. A representative, spectrally dependent
reflectance $R_{sat}(\lambda)$ was constructed as the area-weighted
average of photovoltaic cells (30\%), white thermal control paint
(30\%), multi-layer insulation (30\%) and miscellaneous dark materials
(10\%), using published optical properties \citep{parretta1999angle,
sokolovskiy2022study, palm2012wavelength, bradley2020optical}. 

The
magnitude scale was defined using a blackbody approximation to the
average calibration star, whose effective temperature we derive as
$\sim$7,000~K (F type) by weighting spectral types of Gaia DR3 stars
brighter than $G = +9.0$ by their frequency. Integrating the satellite
and reference spectra over the instrumental and $V$ bandpasses yields a
colour index $(m_{inst} - m_V) = -0.02$~mag, which is effectively zero
given the uncertainties. Hence our raw instrumental magnitudes are
equivalent to $V$-band magnitudes for solar-type reflectors. Sensitivity
tests show the result is robust to the assumed surface composition: even
90\% MLI coverage shifts the index only to $-0.03$~mag, well within the
$\pm0.15$~mag instrumental precision. The full derivation, the adopted
material fractions and their justification, and the resulting spectra
are given in Supplementary Note 2 and Supplementary Fig.~1.

\subsection{Observing campaign and data reduction}

As part of the commissioning phase of the Luciole system, a dedicated
observational campaign was conducted on LEO satellites with the
narrow-field (25mm lens) sensors. The campaign ran for 36 nights, from 7 October to
11 November 2024, at Elginfield Observatory in southwestern Ontario,
Canada ($43.19^{\circ}$N, $81.32^{\circ}$W). Only narrow-field cameras
were used, because of their higher sensitivity. Six cameras operated in
a coordinated, uncued staring mode, each with a FOV of approximately
$17^{\circ} \times 10^{\circ}$ (170.52~deg$^{2}$) and a limiting
sensitivity of $+9.0$~mag in $V$.

Four cameras pointed roughly north, two at elevations of
$\sim$40$^{\circ}$ and two at $\sim$55$^{\circ}$; the lower pair was directed $\sim$14$^{\circ}$ east and
$\sim$12$^{\circ}$ west of due north, and the higher pair $\sim$35$^{\circ}$ east
and $\sim$33$^{\circ}$ west. Of the remaining two, one pointed west and one south,
both at an elevation of about $45^{\circ}$. No camera pointed east.
Actual pointing directions and field sizes are shown in Extended Data
Fig.~\ref{edfig:fov}.

The systems operated continuously during all clear nighttime hours after
nautical twilight, collecting high-cadence (25~Hz) astrometric and
photometric data for all objects with angular velocities below
2~deg~s$^{-1}$ transiting their fields. In total this comprised 407
hours of possible observing time from nautical twilight to nautical
twilight, of which 57\% were available with clear skies, giving a total
survey interval of 230 clear hours. There was additional partial loss
from hardware downtime of individual cameras, which is  accounted
for in the per-camera observing times used for the flux calculation.

To ensure the highest data quality, all collected data were subject to a
rigorous verification process. For each individual satellite track, the
astrometry and photometry were automatically recalibrated using visible
stars within the field of view to correct for pointing drift or
atmospheric changes. These automated calibrations were then manually
inspected and confirmed. Any datasets that did not meet the required
quality standards due to weather, calibration issues or other anomalies
were excluded from the final analysis. A co-added image of all satellite
detections from a single camera on a representative night is shown in
Extended Data Fig.~\ref{edfig:fov}.

Not every analysis uses the full sample of 24,812 tracks. The
magnitude--altitude analyses (Figs~\ref{fig:mag_scatter} and \ref{fig:errbars}
and Extended Data Fig.~\ref{edfig:density}) use the 18,666 tracks of
objects with complete catalogue orbital elements and launch dates. The flux (Extended Data Table~\ref{edtab:flux}) uses the
15,591 passages recorded by all six cameras within the twilight windows.

\subsection{Theoretical passage counts}

Each track is compared with the IAU CPS limit of equation~(\ref{eq:iau})
evaluated at the mean of the object's catalogued perigee and apogee altitudes
(CelesTrak, 2 June 2025). Percentages of non-compliant satellite passes are
computed relative to theoretically possible observations. These are defined as instances in which an
object was predicted from public two-line elements to enter the field of
view of a camera while that camera was operating, after accounting for
cloud cover, camera downtime and invisibility due to atmospheric
attenuation and Earth-shadow entry. This normalisation removes the
dependence on constellation size and is therefore the appropriate
statistic for projecting the effect of a constellation that is scaled up
in number.

\subsection{Global satellite visibility model}

To quantify the potential observation time for a generic constellation
we developed a numerical model that computes the total nightly duration
during which any part of the constellation is visible to a ground-based
observer under specified geometric constraints. The model simplifies the
constellation into a continuous spherical shell of constant altitude,
taken as 550~km, and samples its visibility over a 24-hour period. The
fundamental assumption is that a satellite is visible if it is
illuminated by the Sun and its solar phase angle with respect to the
ground observer lies within a prescribed range. We further require an
apparent elevation of at least $10^{\circ}$ to avoid edge effects, local
obstruction and strong atmospheric extinction. Simulations were
performed and predictions made for satellites brighter than the IAU CPS limit appearing
during the intensive visibility period after sunset and before dawn. The
adopted phase-angle range, 25.9$^{\circ}$--123.2$^{\circ}$, is the
2.5\%--97.5\% percentile interval of the full survey observed set.

The model is established in a geocentric Earth-centred inertial (ECI)
frame with the Earth modelled as a sphere of mean radius $R_E =
6371$~km and the constellation as a concentric shell of radius $R_S =
R_E + h$. Solar ephemerides and the transformation from the observer's
geographic coordinates (latitude $\phi$, longitude $\lambda$) to the ECI
observer vector $\vec{O}$ were computed with the Skyfield library
\citep{rhodes2019skyfield}. Visibility is evaluated on a 5~min time grid.
At each step $t_i$ the shell is sampled at the 14,400 points $\vec{P}$
formed by the 120 latitudes and 120 longitudes of a Fibonacci lattice
(360 of each for Extended Data Fig.~\ref{edfig:funnel})
\citep{gonzalez2010measurement}, and a segment is deemed observable if
at least one point simultaneously satisfies all of the following four conditions.

First, the observer must be in darkness, defined as a solar altitude
below $-12^{\circ}$ (nautical twilight). Second, the point must lie
above the minimum elevation $\epsilon_{min}$,
\begin{equation}
\frac{(\vec{P} - \vec{O}) \cdot \vec{O}}
     {||\vec{P} - \vec{O}|| \cdot ||\vec{O}||} > \sin(\epsilon_{min}).
\end{equation}
Third, the point must be sunlit rather than in the Earth's shadow, which
is modelled as a cylindrical umbra of radius $R_E$ extending away from
the Sun along the geocentric Sun vector $\vec{S}_{geo}$. A point is in
shadow if
\begin{equation}
(\vec{P} \cdot \hat{S}_{geo} < 0)
\quad \text{and} \quad
\sqrt{||\vec{P}||^2 - (\vec{P} \cdot \hat{S}_{geo})^2} < R_E ,
\end{equation}
where $\hat{S}_{geo}$ is the unit vector to the Sun. Note that in
reality the true umbra is conical (owing to the finite angular size of
the Sun), so this approximation slightly overestimates the shadow radius
at satellite altitude. It can therefore only reduce, never inflate, the
predicted fraction of the night with visible satellites, and for the
altitudes considered here (typically below 1,000~km) it does not
materially affect the reported visibility fractions. The bias is
one-sided. Fourth, the solar phase angle
\begin{equation}
\Phi = \arccos\left(
\frac{(\vec{O} - \vec{P}) \cdot (\vec{S}_{geo} - \vec{P})}
     {||\vec{O} - \vec{P}|| \cdot ||\vec{S}_{geo} - \vec{P}||}
\right)
\end{equation}
must satisfy $\Phi_{\text{min}} \le \Phi \le \Phi_{\text{max}}$.

The total observation duration for a night is the summed duration of all
time steps with at least one visible point; the night length is the
summed duration of all steps in which the observer is in darkness. The
model does not explicitly treat earthshine, but it is implicitly
accounted for through the empirically observed phase-angle range. For
the global analysis the simulation was run for every tenth day of a full
year at latitudes from $70^{\circ}$S to $70^{\circ}$N in $4^{\circ}$
increments, at fixed longitude $0^{\circ}$ (which does not affect the
result). This range covers the main landmasses and the most populated
areas while avoiding singularities at the poles, and extends beyond the
$\pm53^{\circ}$ inclination of the main Starlink shells.

The same model gives the fraction of the sky above $10^{\circ}$ elevation in
which some part of the shell is visible. At Elginfield on 20 October 2024 this is
62--64\% at the end of nautical twilight, falling to $\sim$35\% midway through the
evening window. This value ($\sim$60\%) is used to scale the
whole-sky prediction in Table~\ref{tab:predictions}. For another site the model
gives the corresponding fraction.

To test the sensitivity to the adopted shell height we repeated the calculation
for shells at 350, 480 and 1,200~km. At Elginfield on 20 October 2024 the visible
duration is 2.6, 3.6, 4.2 and 9.8~h for shells at 350, 480, 550 and 1,200~km
respectively (of an 11.2~h night). For comparison, at $-30^{\circ}$ latitude it is 2.1--2.2,
3.0--3.3, 3.6--3.7 and 7.0--7.9~h at the March equinox and June solstice. The
concentration of visibility into twilight is therefore a property of the
population below $\sim$600~km, which dominates current numbers; satellites near
1,200~km remain sunlit for most of the night. The fraction of the sky over which
the shell is visible changes much less (49--65\% at the start of the evening
window for 350--1,200~km).

\subsection{Satellite magnitude distribution}

To scale the plane-of-sky satellite flux to an arbitrary limiting sensitivity we
measured the magnitude distribution of observed satellites and derived a
data-driven model. Following the framework developed for natural objects
by \citep{vida2020new} and \citep{vida2022flux}, we assume that the
logarithm of the cumulative number of observations versus magnitude
follows a linear trend defined by the population index
\begin{equation}
r = \frac{N(M_s + 1)}{N(M_s)},
\end{equation}
where $N(M_s)$ is the cumulative number of satellites brighter than
magnitude $M_s$. Restricting the fit to magnitudes between $+2.5$ and
$+5.2$ (the range over which the logarithm of the cumulative distribution is
linear) gives $r = 4.62$, and the effective limiting magnitude (the magnitude to
which the survey is complete) is $+6.15$. 

Because our sample is dominated by Starlink satellites with magnitudes
close to the detection limit, we extend the slope beyond the limiting
magnitude using radar cross-section (RCS) measurements from the
Celestrak catalogue.\footnote{\url{https://celestrak.org/satcat/satcat.csv},
accessed 7 October 2025.} We emphasize that RCS-to-magnitude conversions
are not precise, but at the small sizes of interest they are the best
available brightness proxy. RCS values were converted using the classical relation
\citep{Williams_1966}
\begin{equation}
M = -2.5 \log_{10}(A \cdot \rho \cdot \beta) - 26.7 + 5\log_{10}(d),
\end{equation}
where $A$ is the RCS in m$^{2}$, $\rho$ the reflectivity (assumed 0.2),
$\beta$ the phase angle coefficient (0.08) and $d$ the distance in
metres, taken to correspond to the orbital altitude. A fit to the
derived magnitudes between $+7$ and $+12$ (almost all within the RCS-extrapolated brightness range) gives a much shallower slope
of 1.72. The choice of $\rho$ and $\beta$ does not affect the slope, as
they only scale the RCS. The reference number of satellites in magnitude
bin $M_1$ can then be scaled to another magnitude $M_2$
\citep{vida2022flux} by
\begin{equation}
N(M_2) = N(M_1)\,r^{M_2 - M_1},
\label{eq:mag_scaling}
\end{equation}
evaluated piecewise with
\begin{equation}
r =
\begin{cases}
4.62, & M \leq 6.15 \\
1.72, & M > 6.15 .
\end{cases}
\label{eq:power-law}
\end{equation}
Both the measured cumulative distribution and the RCS-based extension
are shown in Extended Data Fig.~\ref{edfig:distributions}.

\subsection{Plane-of-sky satellite flux}

For a total observation period $T_{obs}$ during which $N_{sat}$ unique
satellite passages are counted, the plane-of-sky flux is obtained by
normalising for both time and the camera's solid angle $\Omega_{deg}$ in
square degrees, calculated as the product of the horizontal and vertical
angular widths ($FOV_w \times FOV_h$):
\begin{equation}
F_{sat} = \frac{N_{sat}}{T_{obs} \times \Omega_{deg}} .
\label{eq:flux}
\end{equation}
We propose this as a fundamental measure of the apparent surface density
of satellites: it is independent of specific camera hardware or exposure
settings and so permits direct comparison between instruments and sites.
Flux values were computed separately for each camera over the twilight
visibility periods, defined as the two hours after local sunset and the two hours before local
sunrise at Elginfield (of which the portion after the end of nautical twilight
is $\sim$75~min). The observing time of each camera within these windows was
taken as the time spanned by its detections, excluding any gap of 30~min or
longer between successive detections (less 15~min at either end of the gap),
which we treat as cloud or camera downtime. Note that the flux is a direct count and involves no assumption about satellite
altitude. The fixed shell altitude of 550~km enters only the visibility model
used to generalise the timing and sky coverage (see `Global satellite visibility model'). Per-camera survey results are given in Extended
Data Table~\ref{edtab:flux}.

\subsection{Predicting satellite counts for arbitrary observations}

The flux can be used to predict the number of satellites $N_{pred}$
visible in any solid angle of sky from any location and time, expressed
through an effective collection time $T_{eff}$ that accounts for the two
observational regimes. The parameter which distinguishes them is the
average satellite passage time,
\begin{equation}
t_{pass} \approx \frac{\theta_{pass}}{\omega_{sat}},
\end{equation}
the duration a satellite takes to cross the field if it starts at a
random location within it, where $\theta_{pass}$ is the average observed
angular arc in the field and $\omega_{sat}$ the typical angular
velocity. A numerical simulation in which satellites start at random
locations and random headings shows that approximating the
characteristic arc as half the FOV diagonal,
\begin{equation}
\theta_{pass} = \frac{\sqrt{(FOV_w)^2 + (FOV_h)^2}}{2},
\end{equation}
is accurate to within 10\%. From our survey the median angular speed is
$\omega_{sat} = 0.56^{\circ}$~s$^{-1}$, with a 5\%--95\% interval of
0.17--2.76~deg~s$^{-1}$ (Extended Data
Fig.~\ref{edfig:distributions}). The effective collection time is then
\begin{equation}
T_{eff} = t_{exp} + t_{pass},
\end{equation}
and the predicted count for any observation is
\begin{equation}
N_{pred} = F_{sat} \times \Omega_{deg} \times T_{eff} .
\label{eq:main_n_pred}
\end{equation}
In the snapshot regime ($t_{exp} < t_{pass}$) the count is effectively
set by satellites already present in the field, since new ones have
little time to enter, so the observation captures the instantaneous
satellite density; in the flux regime ($t_{exp} \gg t_{pass}$) the
observation is long enough to accumulate new transits and the count is
dominated by satellites entering during the exposure. The full
derivation of equation~(\ref{eq:main_n_pred}) from the ambient satellite
density is given in Supplementary Note 4.

\subsection{Trailing loss and the effective limiting magnitude}

Detection of a satellite by a sidereally tracking telescope is
fundamentally limited by the surface brightness of its trailed streak:
as the satellite moves, its flux is spread across a line of pixels,
reducing the signal-to-noise ratio relative to a stationary source of
the same magnitude. The number of pixels covered during an exposure
$t_{exp}$ on a detector of pixel scale $\theta_{pix}$ is
\begin{equation}
N_{pix} = \frac{\omega_{sat} \times t_{exp}}{\theta_{pix}},
\end{equation}
and the corresponding magnitude loss is
\begin{equation}
\Delta m = 2.5 \log_{10}\left(\frac{N_{pix}}{\mathrm{FWHM}}\right),
\end{equation}
where FWHM is the full width at half maximum, assumed to be 3 pixels for
a well-focused sensor \citep{xin2018study, vida2022flux}. For the Vera
C. Rubin Observatory, with $t_{exp} = 30$~s and $\theta_{pix} =
0.2$~arcsec per pixel \citep{sebag2020vera}, a satellite moving at the
median observed speed produces a streak covering approximately 302,400
pixels and a trailing loss of $\Delta m \approx 12.51$~mag. Given a
nominal stationary limiting magnitude of $+24$ \citep{lochner2022impact},
the effective limiting magnitude for a typical LEO satellite is
\begin{equation}
m_{lim,sat} = m_{lim,stationary} - \Delta m \approx 24 - 12.51 = 11.49 .
\end{equation}
Scaling the measured flux to this limit using
equations~(\ref{eq:mag_scaling}) and (\ref{eq:power-law}) gives an
enhancement factor relative to the IAU CPS limit of $1.72^{11.49-7.0}
= 11.4$. Applying the same analysis to a
consumer-grade APS-C camera with a 20~mm f/2.8 lens and 5~$\mu$m pixels
(image scale $\sim$51.6~arcsec per pixel) gives a streak of $\sim$391
pixels in a 10~s exposure, $\Delta m \approx 5.3$~mag, and, for a
measured stellar limiting magnitude of $+8.0$, an effective satellite
limiting magnitude of $+2.71$. For the naked-eye scenario we adopt an
effective field of view of order 7,000~deg$^{2}$ for an observer looking
upward at $45^{\circ}$ \citep{rendtel_handbook_2022}, and, for the whole-sky scenario only, scale the prediction to the $\sim$60\%
of the sky over which satellites are visible at our latitude (see `Global
satellite visibility model').

\section*{Data availability}

The satellite detection and photometry data underlying the analyses,
figures and tables in this paper are available from the corresponding
author upon reasonable request.

\section*{Code availability}

The core code used for this analysis is part of our research group's
GitHub organisation and can be found at \url{https://github.com/wmpg/}.

\section*{Corresponding author}

Correspondence and requests for materials should be addressed to Denis
Vida (dvida@uwo.ca).

\section*{Acknowledgements}

The authors gratefully acknowledge funding from the Rotman Institute of
Philosophy's Think Tank program and from Defence Research and
Development Canada under contract W7714-238467/001/sl. This work was
made possible by funding from the Canadian Natural Sciences and
Engineering Research Council (NSERC) Alliance program (ALLREP
586010-23) and the NSERC Department of National Defence Supplement
program (RGPIN 2023-03538).

\section*{Author contributions}

J.R. --- method development, visualisations, manuscript writing. D.V. ---
project conceptualisation, project leadership, method development,
manuscript writing. P.B. --- funding acquisition, supervision, manuscript
writing. P.G.B. --- funding acquisition, project administration,
supervision, manuscript writing, analysis and editing. S.G. --- project support,
manuscript editing. M.J.M. --- hardware development, observational
methods. D.L.C. --- software development. M.V. --- data analysis support.
S.M. --- supervision, project support. L.S. --- funding acquisition,
supervision, technical advising.

\section*{Competing interests}

The authors declare no competing interests.

\section*{Additional information}

\noindent\textbf{Supplementary information} is available for this paper
(Supplementary Notes 1--5, Supplementary Figs.~1--7 and Supplementary
Table~1).

\clearpage

\clearpage

\begin{table}[p]
\centering
\caption{\textbf{Predicted number of satellites for representative
observing scenarios.} Values are computed from the combined measured
flux of $3.51 \times 10^{-3}$ satellites deg$^{-2}$ min$^{-1}$ at the effective completeness limit of $+6.15$, scaled to $5.56 \times 10^{-3}$ satellites deg$^{-2}$ min$^{-1}$ brighter than the IAU CPS limit of $+7.0$ with equation~(\ref{eq:power-law}), using
equation~(\ref{eq:main_n_pred}). $N(+7)$ is the predicted number
brighter than the IAU CPS limit; $N$(LM) is the predicted number above
the effective limiting magnitude of the instrument for a moving target,
after trailing loss. $\dagger$Visibility simulations show that
satellites can be observed over at most $\sim$60\% of the sky at our
study latitude ($+43^{\circ}$); the flux has been scaled to 60\% of the
predicted value to reflect this.}
\label{tab:predictions}
\begin{tabular}{lcccccc}
\toprule
Scenario & Lim.\ mag. & FOV & $\Omega_{deg}$ (deg$^2$) & $t_{exp}$ (s) & $N(+7)$ & $N$(LM) \\
\midrule
Human vision                & $+6.5$  & 50$^{\circ}$ radius & 7,000  & 0.04 & $\sim$58.0 & $\sim$44.2 \\
Whole sky$^{\dagger}$       & $+6.5$  & hemisphere          & 20,627 & 0.04 & $\sim$184.4 & $\sim$140.6 \\
APS-C camera (20~mm)        & $+2.7$  & 60$^{\circ} \times$ 40$^{\circ}$ & 2,400 & 10 & $\sim$16.6 & $\sim$0.054 \\
Vera C. Rubin Observatory   & $+11.5$ & 3.5$^{\circ}$ diam. & 9.62   & 30 & $\sim$0.030 & $\sim$0.34 \\
\bottomrule
\end{tabular}
\end{table}

\clearpage

\setcounter{figure}{0}
\setcounter{table}{0}
\renewcommand{\thefigure}{\arabic{figure}}
\renewcommand{\thetable}{\arabic{table}}
\renewcommand{\figurename}{Extended Data Fig.}
\renewcommand{\tablename}{Extended Data Table}

\section*{Extended Data}

\begin{figure}[p]
\centering
\includegraphics[width=\linewidth]{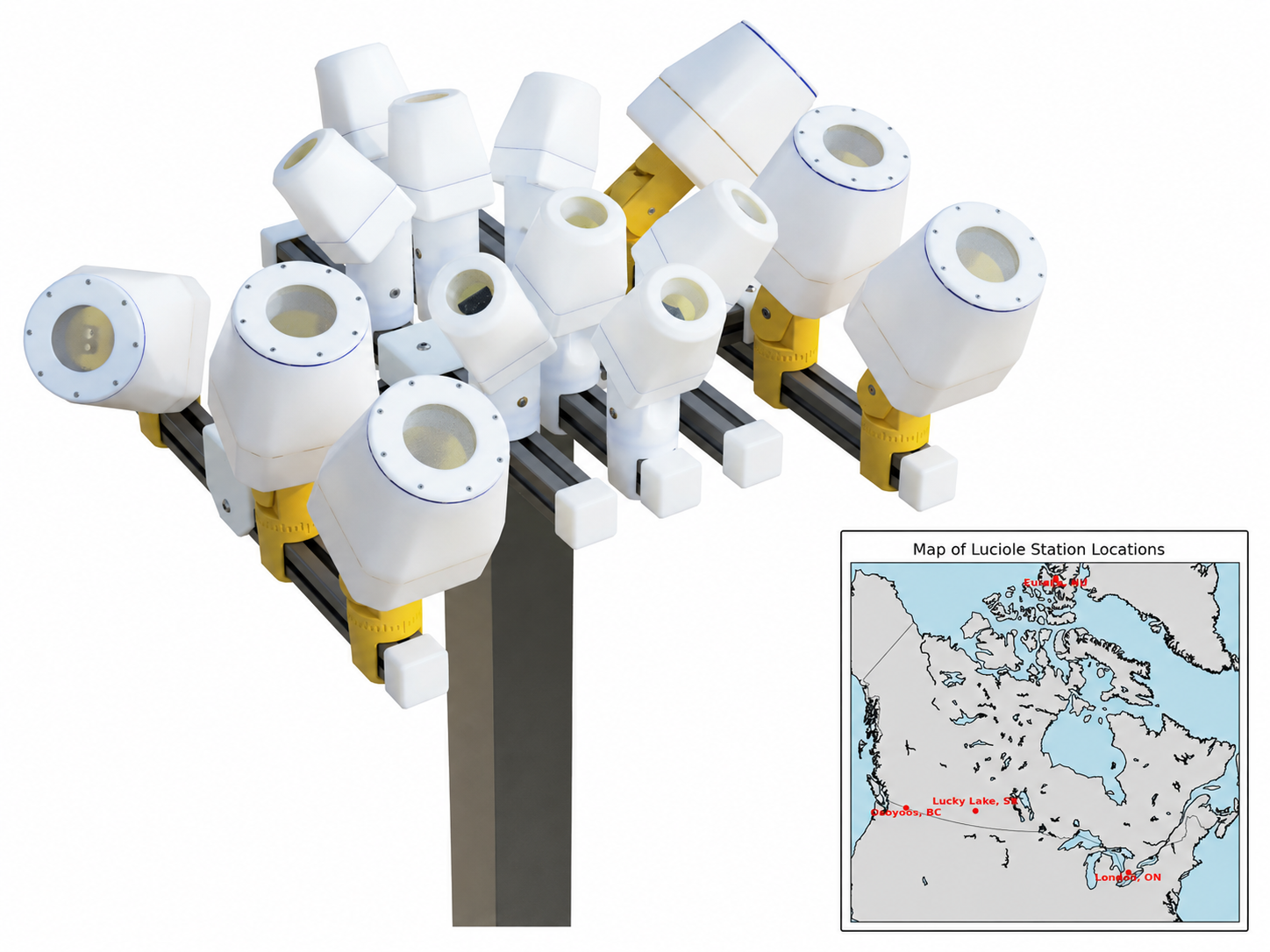}
\caption{\textbf{The Luciole camera system.} A render of the Luciole
``fly's eye'' array. Wide-field 8~mm cameras occupy the middle of the
array while the more sensitive narrow-field 25~mm cameras are placed at
the edges. The inset shows the Luciole camera sites currently active in
Canada (red labelling), in Ontario, Saskatchewan, British Columbia and
Nunavut. Only the six narrow-field cameras at the Elginfield, Ontario
site were used for the survey reported here.}
\label{edfig:system}
\end{figure}

\begin{figure}[p]
\centering
\includegraphics[width=0.92\linewidth]{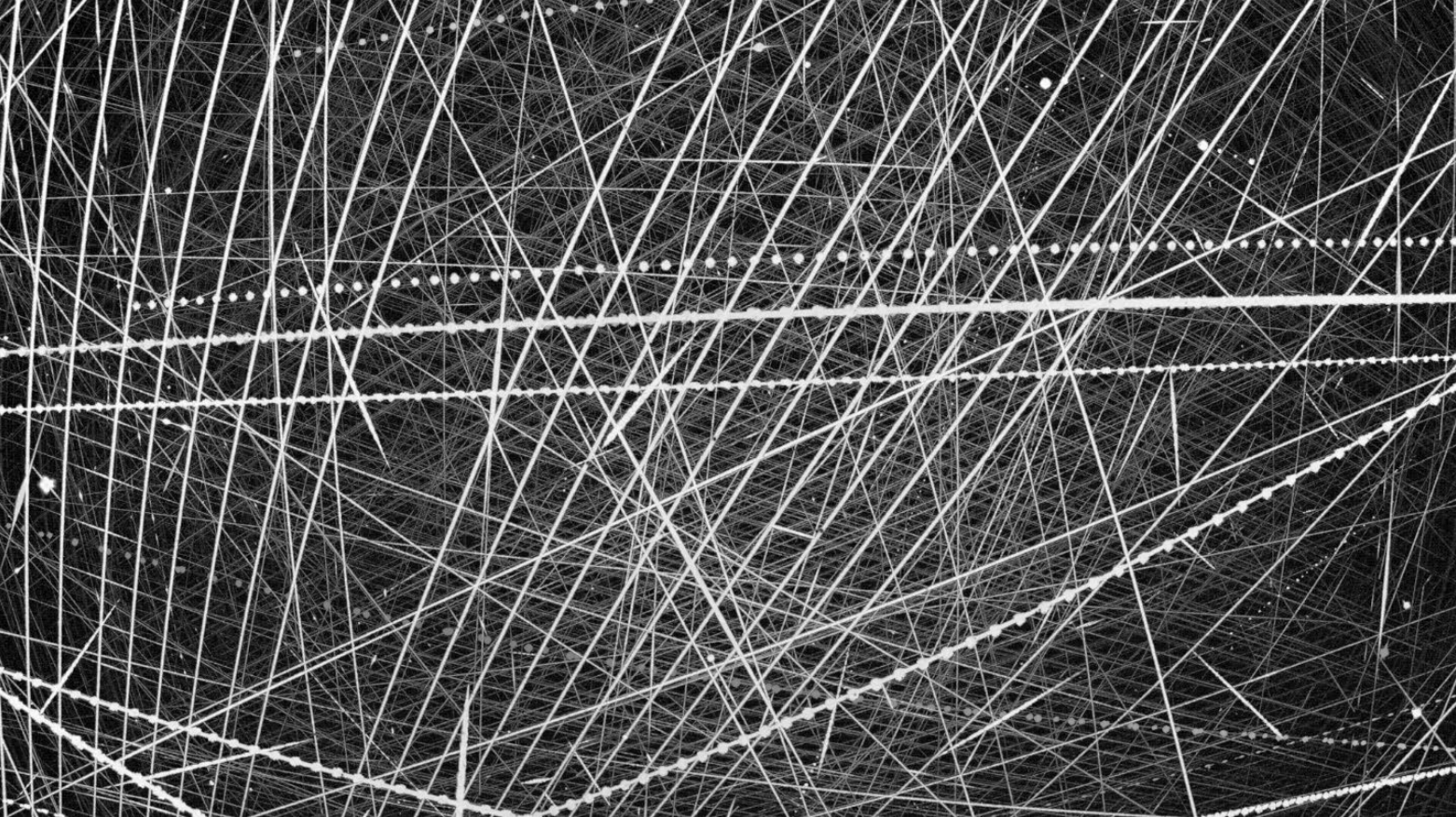}\\[2ex]
\includegraphics[width=0.72\linewidth]{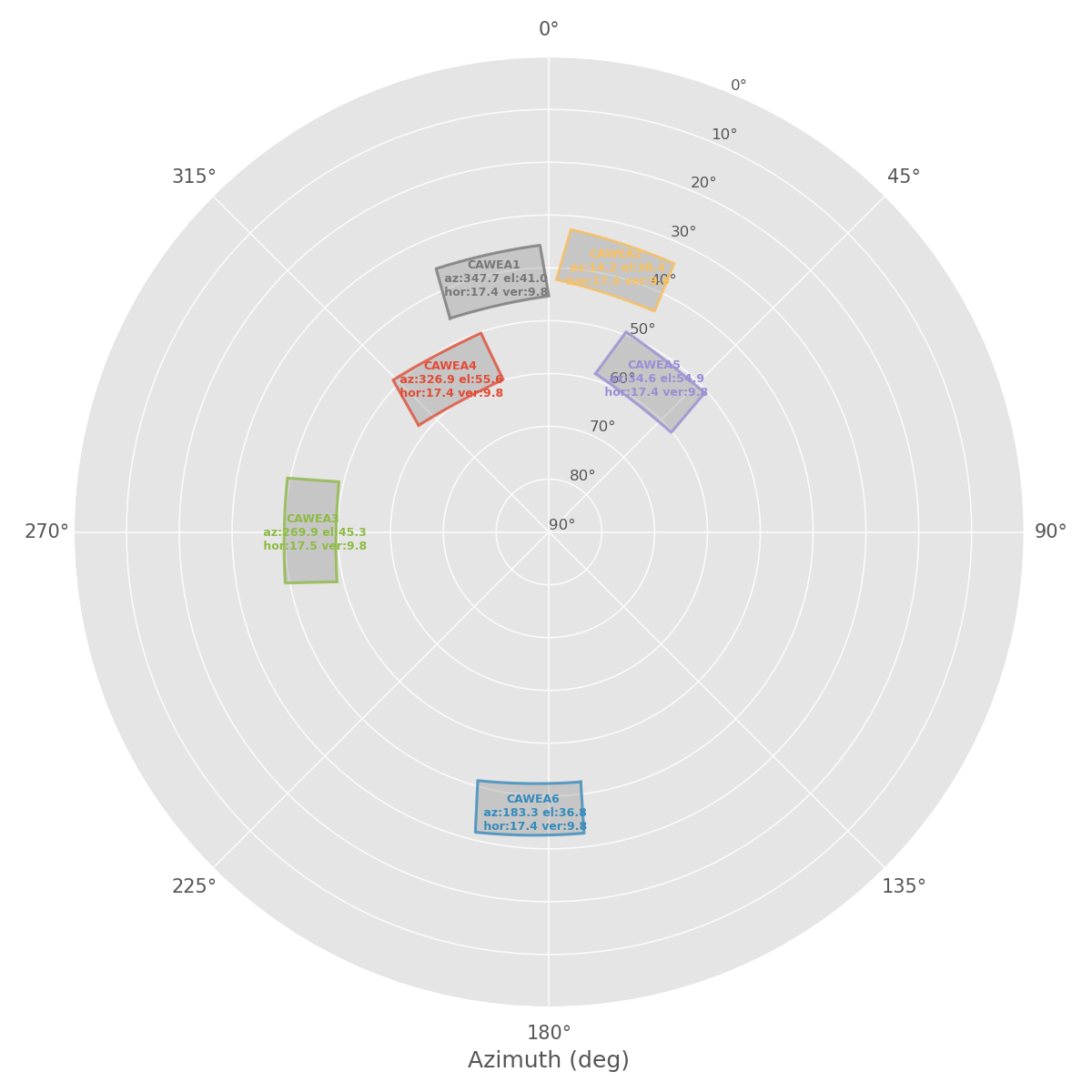}
\caption{\textbf{Survey data and sky coverage.} \textbf{a}, Co-added
image of all satellite detections on the night of 31 May--1 June 2024
from a single camera. Objects appearing as dotted lines are aircraft
strobes and are rejected by software; meteor trails and trailed star
images are also visible. \textbf{b}, Hemispherical plot of the sky as
seen from the Elginfield station showing the field of view of each 25~mm
narrow-field Luciole camera used in the survey. Grey, camera 1; yellow,
camera 2; green, camera 3; red, camera 4; purple, camera 5; blue, camera
6. No camera pointed east, which produces the sunset/sunrise asymmetry
in Fig.~\ref{fig:timing}a.}
\label{edfig:fov}
\end{figure}

\begin{figure}[p]
\centering
\includegraphics[width=\linewidth,trim={3.5cm 0 6cm 0cm},clip]{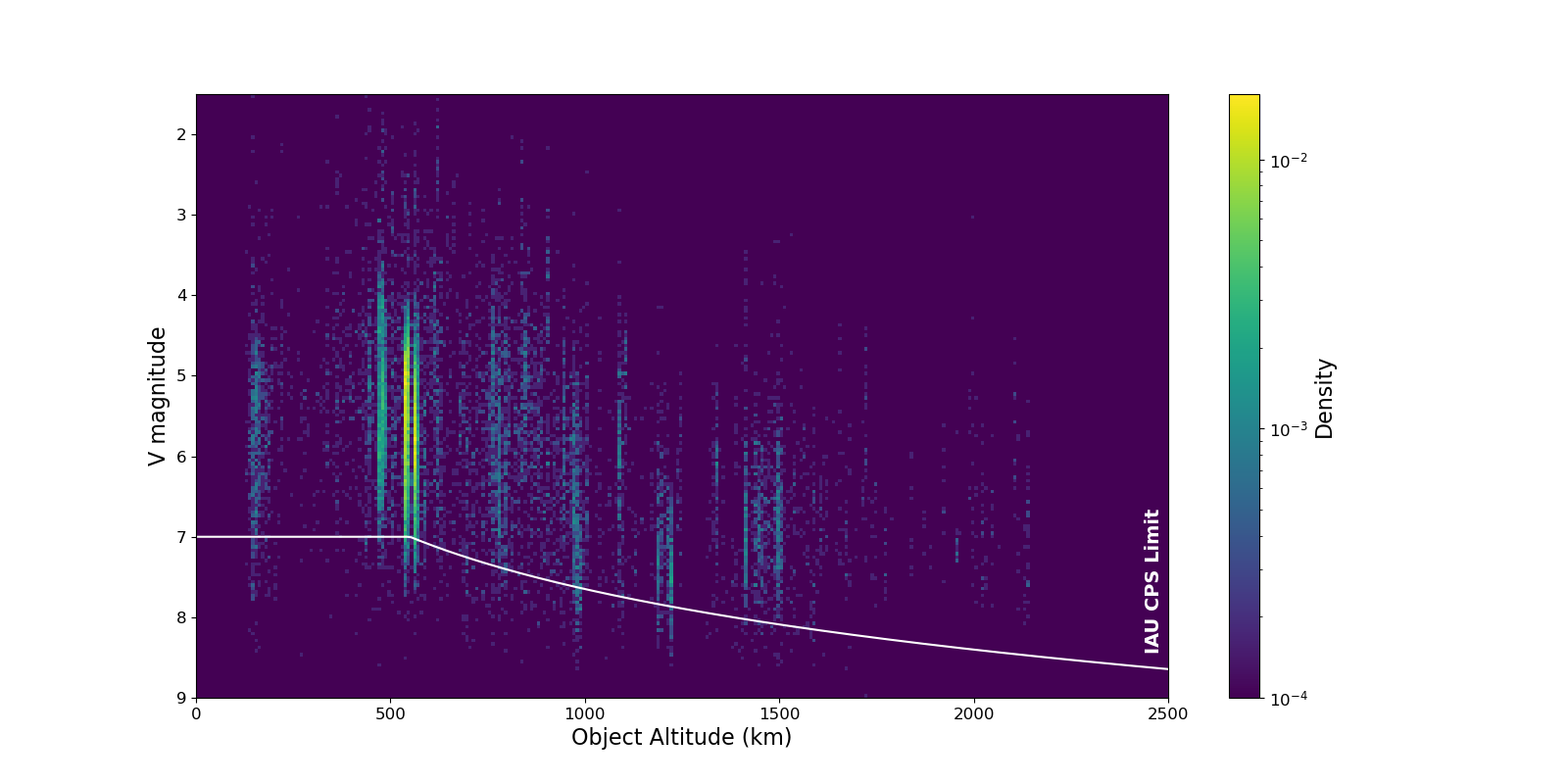}
\caption{\textbf{Density of observed satellite tracks in magnitude and
altitude.} Relative density (colour bar) of satellite tracks observed by
Project Luciole during October--November 2024 from Elginfield, Ontario,
Canada. Magnitude is the median visual magnitude of the track points and
altitude is that of the observed object. The white line is the IAU CPS
brightness limit. The highest concentration of tracks corresponds to the
dense Starlink shell near 550~km and lies almost entirely on the
non-compliant side of the limit, showing that a typical satellite at this
altitude is overwhelmingly likely to appear brighter than the IAU
recommendation.}
\label{edfig:density}
\end{figure}

\begin{figure}[p]
\centering
\includegraphics[width=\linewidth]{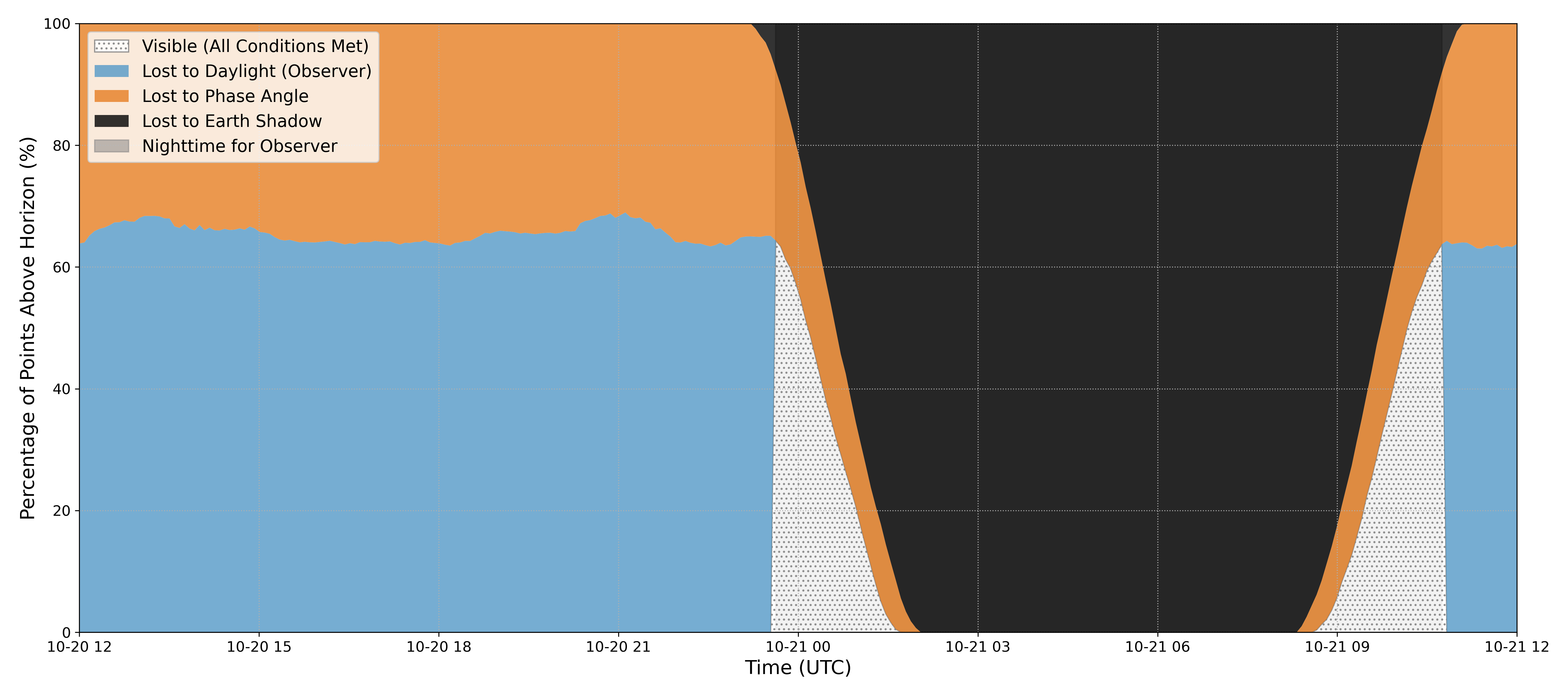}
\caption{\textbf{Successive geometric filters determining satellite
visibility.} The visibility funnel for a 550~km satellite shell,
simulated for Elginfield Observatory on 20--21 October 2024, about
halfway through the data collection campaign, using a minimum elevation
of 10$^{\circ}$ and a phase-angle range of
25.9$^{\circ}$--123.2$^{\circ}$. The total height of the coloured area
(100\%) at any time represents all sample points above the observer's
horizon. This set is first reduced by points in the Earth's shadow
(black). The remaining sunlit points are then filtered by the required
solar phase angle range (orange). Finally, true visibility requires the
observer to be in darkness (grey shaded background); points meeting all
other criteria during local daylight are discarded (light blue). The
final visible population is the hatched white area within the nighttime
window, totalling 4~h~20~min on this date. Time is given as month-day and
UT hour.}
\label{edfig:funnel}
\end{figure}

\begin{figure}[p]
\centering
\includegraphics[width=\linewidth]{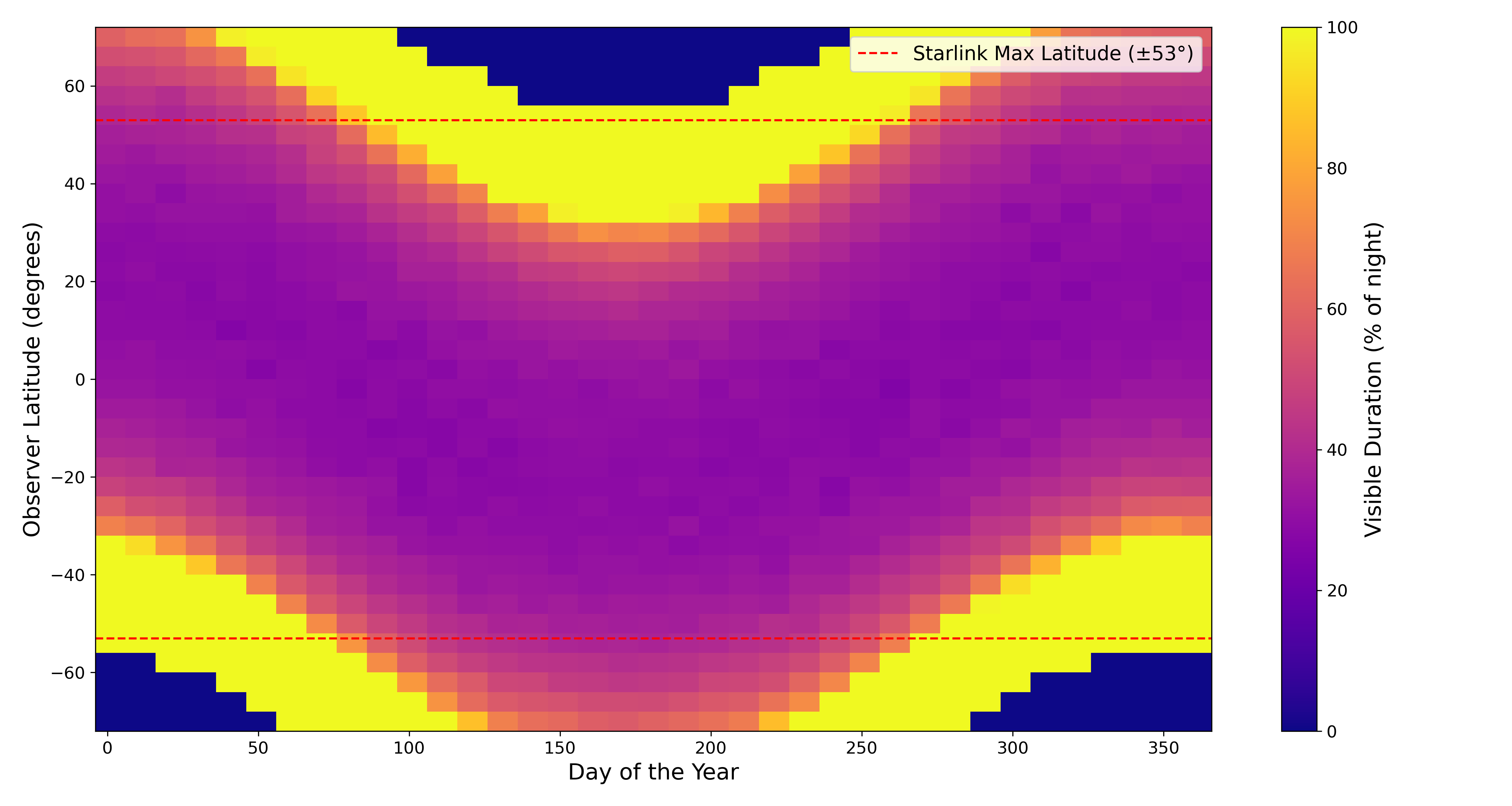}
\caption{\textbf{Satellite visibility as a fraction of the night.}
Percentage of the night (colour bar) during which some part of a 550~km
satellite shell is visible, as a function of day of year and observer
latitude. Horizontal red lines mark the latitudinal extent of the
Starlink megaconstellation, which dominates the visibility. Dark areas
show the latitudes and times of year at which the Earth's surface has
constant daylight. The greatest fractional impact occurs in summer at
high latitudes, where satellites remain illuminated for the entire
night, although the absolute observing time lost is then small because
the nights are short.}
\label{edfig:vispct}
\end{figure}

\begin{figure}[p]
\centering
\includegraphics[width=0.86\linewidth]{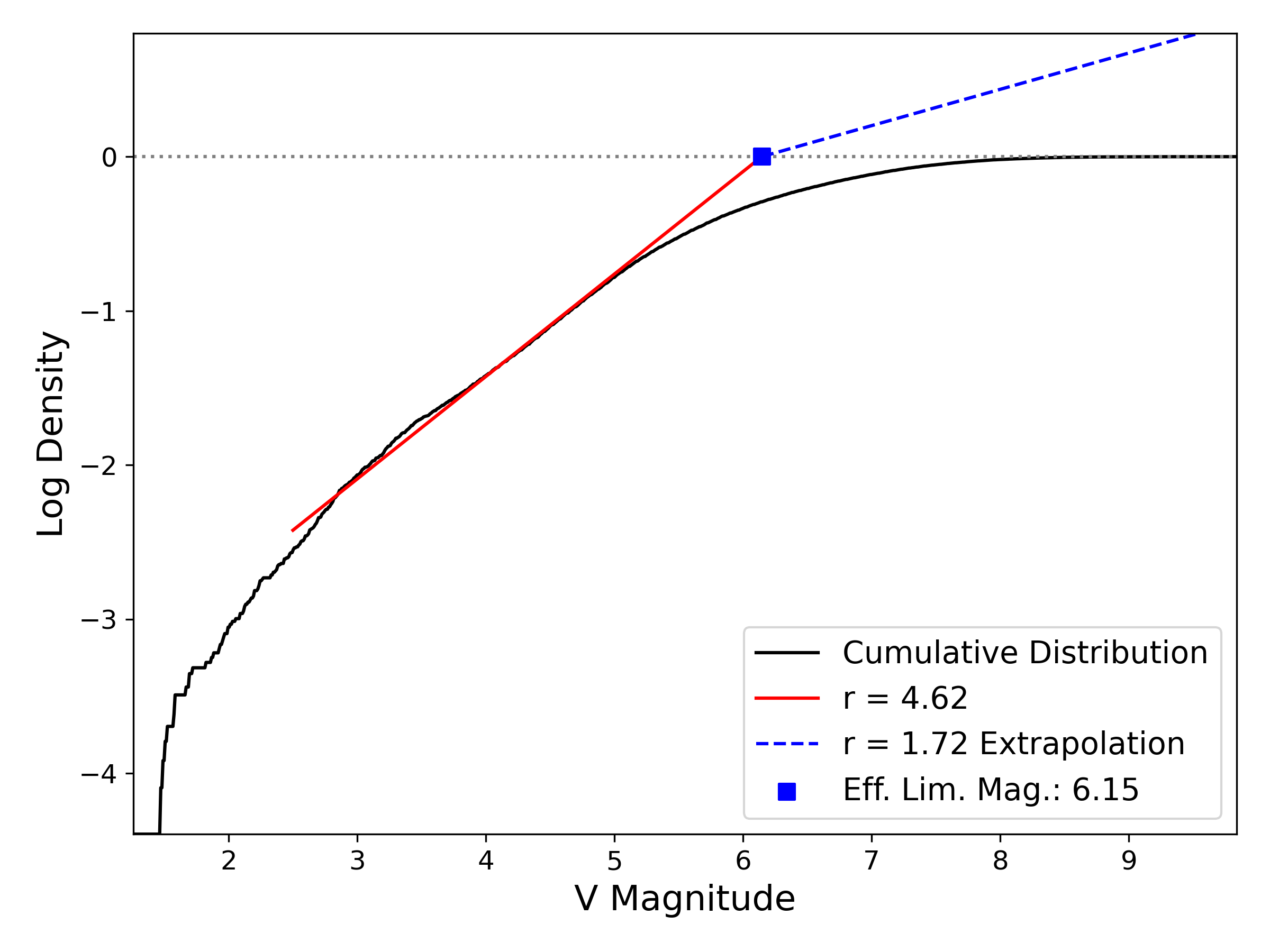}\\[2ex]
\includegraphics[width=0.86\linewidth,trim={0cm 0 0cm 0},clip]{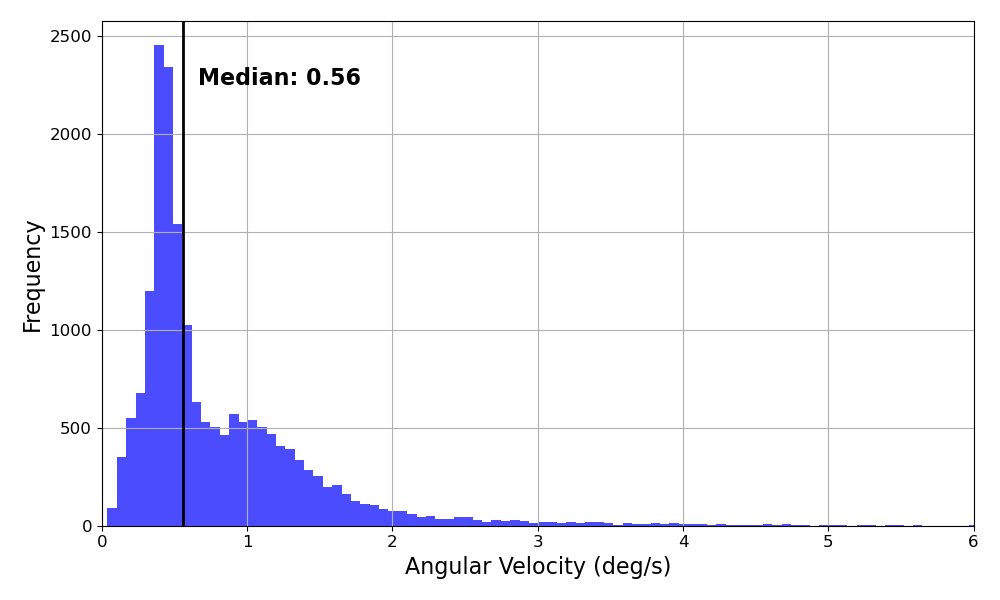}
\caption{\textbf{Brightness and angular speed distributions of the
observed sample.} \textbf{a}, Base-10 logarithm of the cumulative number
of satellites observed by the survey brighter than a given $V$ magnitude
(solid black curve). The red line is the adopted satellite brightness
distribution model and the blue square marks the effective limiting
magnitude of the survey observed sample, $+6.15$ using the assumptions of our piecewise power-law model. The blue dashed line is the
number distribution extrapolated to fainter magnitudes from catalogued
radar cross-section measurements. The two segments define the population
indices $r=4.62$ and $r=1.72$ used in equation~(\ref{eq:power-law}).
\textbf{b}, Histogram of angular velocities for all observed satellite
passages; each passage contributes its own median value. The overall
median of 0.56~deg~s$^{-1}$ is shown as a black line, and the 5\%--95\%
percentile interval is 0.17--2.76~deg~s$^{-1}$.}
\label{edfig:distributions}
\end{figure}

\begin{figure}[p]
\centering
\includegraphics[width=\linewidth]{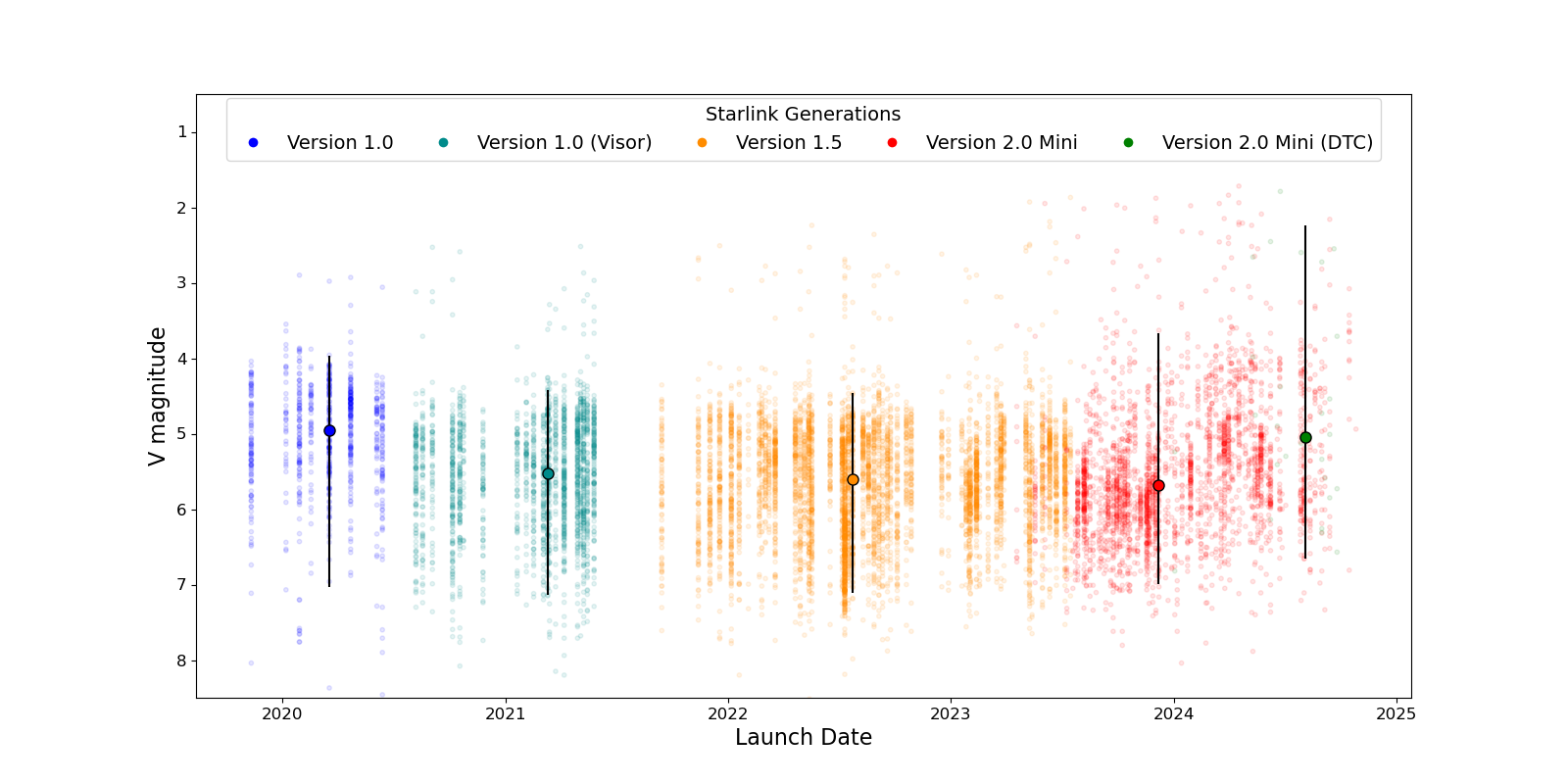}
\caption{\textbf{Evolution of Starlink brightness with generation and
launch date.} Visual magnitude of observed Starlink passages plotted
against the launch date and generation of the satellite as of the time of our survey in fall 2024. The plot shows
the initial dimming trend for satellite launched from 2020 through early 2024 following the
introduction of the VisorSat and later designs, and the subsequent sharp
increase in brightness with the launch of the Version 2.0 Mini
Direct-to-Cell variants, which return close to the brightness of the
original Version 1.0 spacecraft. Large points give the median magnitude of each generation, plotted at its median
launch date, and bars the 2.5\%--97.5\% percentile interval. Phase-angle-resolved
plots for each individual generation are given in Supplementary Figs.~2--6.}
\label{edfig:starlinkgens}
\end{figure}

\begin{figure}[p]
\centering
\includegraphics[width=\linewidth]{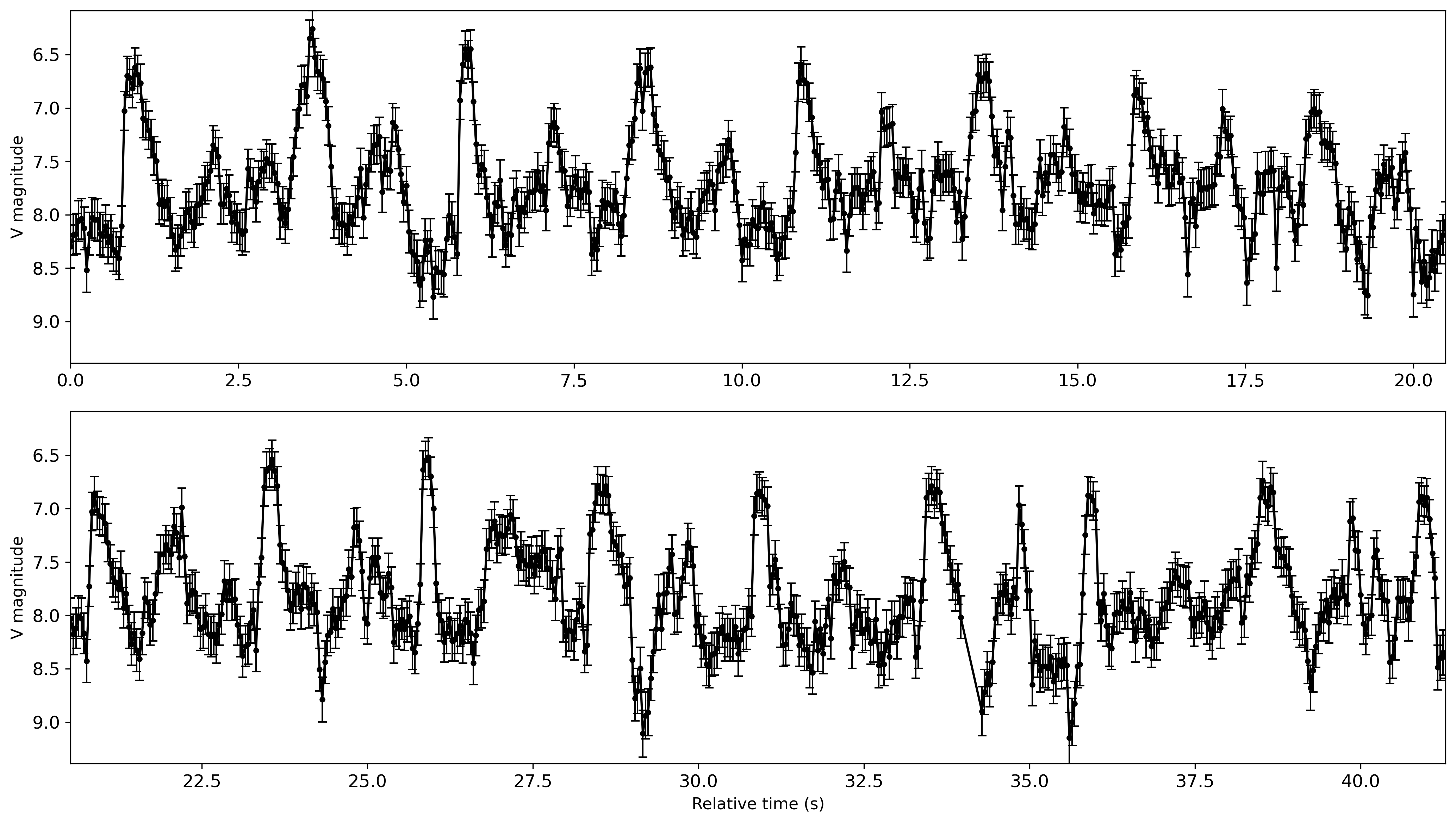}
\caption{\textbf{High-cadence lightcurve of the tumbling satellite
Cosmos 2553.} Measured lightcurve of Cosmos 2553 observed on 3 October
2025 starting at 07:54:47~UTC. Error bars show $1\sigma$ photometric
uncertainties. The 25~Hz photometry resolves a rotation period of
$\sim$1.25~s and periodic brightness flares that vary by about 2.5
magnitudes, demonstrating the system's ability to characterise satellite
attitude and stability.}
\label{edfig:cosmos}
\end{figure}

\clearpage

\begin{sidewaystable}
\centering
\caption[Brightness of satellite constellations and their adherence to
the IAU CPS brightness limit]{\textbf{Brightness of satellite
constellations and their
adherence to the IAU CPS brightness limit.} Entries are ranked by the
number of observed passages brighter than the IAU CPS limit. Most
categories are either a satellite constellation (e.g.\ STARLINK) or a
class of debris (e.g.\ SL, denoting rocket bodies from Soviet or Russian
launch vehicles). Object counts are taken from Celestrak on 2 June 2025
and so differ slightly from the population observed at the end of 2024.
MC(2025) gives apparent $V$-band magnitudes for the constellation where
available from \protect\citep{mallama_satellite_2025}. $^*$Starlink
version
counts are approximate, calculated from the number deployed according to
the only open-source information available (Starlink launches,
\protect\url{https://en.wikipedia.org/wiki/List_of_Starlink_and_Starshield_launches},
accessed 1 September 2025).
$^{\dagger}$The rocket-body row aggregates every rocket-body track regardless of
launcher, and so overlaps the launcher-family rows (SL, CZ, BREEZE, DELTA, THOR and
ATLAS), which also include other debris from those launchers. $^{\dagger\dagger}$The Starlink v2.0 Mini
category includes both Direct-to-Cell and internet communication
variants, which often have different magnitudes.}
\label{edtab:constellations}
\footnotesize
\begin{tabular}{clrrrrrrr}
\toprule
Rank & Constellation/object type & Total objects & Passages & Brighter
than limit & \% brighter & Median $V$ & $V$ 2.5--97.5\% & MC(2025) \\
\midrule
1  & STARLINK                          & 8722  & 7422 & 7074 & 95  & 5.6 & 4.1---7.2 &  \\
2  & STARLINK v1.5                     & 2875$^*$ & 3976 & 3768 & 95  & 5.7 & 4.4---7.2 & 5.67 \\
3  & Rocket bodies$^{\dagger}$         & 6752  & 1953 & 1869 & 96  & 5.9 & 3.2---8.0 &  \\
4  & STARLINK v1.0 Visor               & 1187$^*$ & 1775 & 1702 & 96  & 5.5 & 4.5---7.1 & 6.43 \\
5  & SL                                & 3789  & 1141 & 1080 & 95  & 6.1 & 3.6---8.1 &  \\
6  & STARLINK v2.0 Mini$^{\dagger\dagger}$ & 1586$^*$ & 1023 & 974 & 95 & 5.8 & 3.6---7.1 & 5.2--6.2 \\
7  & COSMOS                            & 2584  & 1004 & 899  & 90  & 6.7 & 3.6---8.1 &  \\
8  & STARLINK v1.0                     & 478$^*$  & 621  & 603  & 97  & 5.0 & 4.0---7.2 & 5.05 \\
9  & GLOBALSTAR                        & 85    & 462  & 457  & 99  & 6.7 & 4.5---7.9 &  \\
10 & CZ                                & 500   & 457  & 443  & 97  & 5.5 & 2.9---7.7 &  \\
11 & ONEWEB                            & 656   & 300  & 255  & 85  & 7.3 & 6.3---8.3 & 7.85 \\
12 & YAOGAN                            & 162   & 194  & 188  & 97  & 5.9 & 4.0---7.7 &  \\
13 & IRIDIUM                           & 175   & 181  & 178  & 98  & 6.2 & 4.4---7.3 &  \\
14 & BREEZE                            & 120   & 146  & 145  & 99  & 5.9 & 3.6---7.8 &  \\
15 & METEOR                            & 64    & 139  & 134  & 96  & 5.8 & 3.5---7.6 &  \\
16 & DELTA                             & 476   & 128  & 123  & 96  & 6.2 & 3.6---8.1 &  \\
17 & GONETS                            & 33    & 100  & 99   & 99  & 6.8 & 5.3---7.9 &  \\
18 & THOR                              & 90    & 101  & 93   & 92  & 6.2 & 4.4---8.1 &  \\
19 & GEESAT                            & 30    & 102  & 91   & 89  & 6.3 & 5.1---7.4 &  \\
20 & OBJECT                            & 605   & 85   & 80   & 94  & 5.8 & 3.7---7.7 &  \\
21 & OPS                               & 412   & 82   & 78   & 95  & 6.2 & 4.4---8.3 &  \\
22 & YUNHAI                            & 18    & 75   & 68   & 91  & 6.3 & 4.6---7.8 &  \\
23 & ORBCOMM                           & 60    & 71   & 63   & 89  & 6.3 & 4.5---7.8 &  \\
24 & DMSP                              & 12    & 62   & 59   & 95  & 6.1 & 4.7---7.6 &  \\
25 & MIDAS                             & 6     & 52   & 52   & 100 & 7.4 & 6.4---8.3 &  \\
\multicolumn{9}{c}{\dots} \\
28 & FENGYUN                           & 29    & 38   & 37   & 97  & 5.1 & 3.7---7.5 &  \\
\multicolumn{9}{c}{\dots} \\
49 & GUOWANG/HULIANWANG                & 34    & 18   & 18   & 100 & 4.8 & 4.3---5.8 &  \\
\bottomrule
\end{tabular}
\normalsize
\end{sidewaystable}

\begin{sidewaystable}[p]
\centering
\caption{\textbf{Measured plane-of-sky satellite flux for each survey
camera.} Flux is computed from equation~(\ref{eq:flux}) using a uniform
per-camera field of view of 170.52~deg$^2$. The pointing azimuth and
elevation give the centre of the field. Observed passages are those
recorded during the twilight visibility periods; observing times exclude cloud and per-camera downtime, identified as
gaps of 30~min or more between detections (Methods). Flux is given both for
all detected objects and for the subset brighter than the IAU CPS limit.
The factor of three spread between cameras reflects their different
pointing directions and demonstrates the anisotropy of the satellite
distribution on the sky.}
\label{edtab:flux}
\begin{tabular}{rrrrrrrr}
\toprule
Camera & Azim. & Elev. & Observed & Brighter than & Obs.\ time & Flux (all) & Flux ($>$ IAU limit) \\
       & (deg) & (deg) & passages & IAU limit     & (h)        & (sats deg$^{-2}$ min$^{-1}$) & (sats deg$^{-2}$ min$^{-1}$) \\
\midrule
1 & 348 & 41 & 3,879 & 3,739 & 92.17 & $4.11 \times 10^{-3}$ & $3.96 \times 10^{-3}$ \\
2 & 14  & 38 & 3,486 & 3,249 & 76.15 & $4.47 \times 10^{-3}$ & $4.17 \times 10^{-3}$ \\
3 & 270 & 45 & 3,887 & 3,848 & 68.23 & $5.56 \times 10^{-3}$ & $5.51 \times 10^{-3}$ \\
4 & 327 & 56 & 1,376 & 1,308 & 79.31 & $1.69 \times 10^{-3}$ & $1.61 \times 10^{-3}$ \\
5 & 35  & 55 & 1,588 & 1,527 & 73.67 & $2.10 \times 10^{-3}$ & $2.03 \times 10^{-3}$ \\
6 & 183 & 37 & 1,375 & 1,326 & 44.98 & $2.98 \times 10^{-3}$ & $2.88 \times 10^{-3}$ \\
\midrule
\textbf{Total} & & & 15,591 & 14,997 & 434.51 & $\mathbf{3.51 \times 10^{-3}}$ & $\mathbf{3.37 \times 10^{-3}}$ \\
\bottomrule
\end{tabular}
\end{sidewaystable}

\end{document}